\documentclass{WileyMSP-template}

\usepackage{ulem}
\usepackage{amsmath}
\usepackage{mathtools,cases}
\usepackage{stfloats}
\usepackage{xcolor}
\usepackage{float}
\usepackage{url}
\usepackage{graphicx}
\usepackage{dcolumn}
\usepackage{bm}
\usepackage{tabularx}
\usepackage{multirow}
\usepackage{ragged2e}
\usepackage{lineno}

\begin{document}

\pagestyle{fancy}
\rhead{\includegraphics[width=1.5cm]{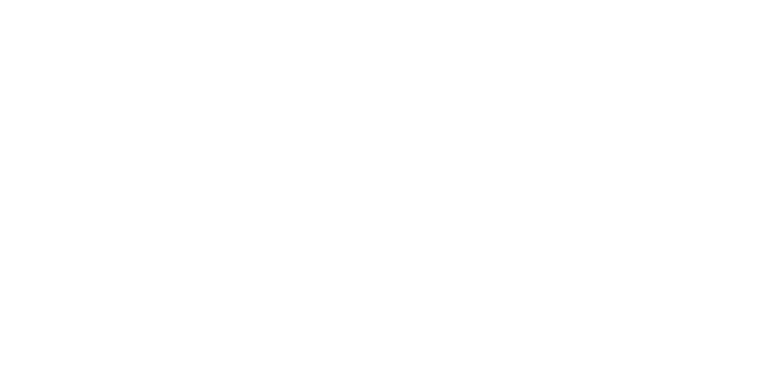}}

\title{Dipole-quadrupole model and multipole analysis of \\ resonant membrane metasurfaces}

\maketitle

\author{Izzatjon Allayarov$^{1,2,3,*}$}
\author{Andrey B. Evlyukhin$^{3,4}$}
\author{Antonio Cal{\`a} Lesina$^{1,2,3}$}

\begin{affiliations}
$^{1}$Hannover Centre for Optical Technologies, Leibniz University Hannover, Hannover, Germany\\
$^{2}$Institute of Transport and Automation Technology, Leibniz University Hannover, Garbsen, Germany\\
$^{3}$Cluster of Excellence PhoenixD, Leibniz University Hannover, Hannover, Germany\\
$^{4}$Institute of Quantum Optics, Leibniz University Hannover, Hannover, Germany\\
$^{*}$izzatjon.allayarov@hot.uni-hannover.de

\end{affiliations}


\keywords{Membrane metasurface, Coupled dipole-quadrupole approach, Multipole decomposition}

\justifying
\begin{abstract}

Membrane-metasurfaces, formed by periodic arrangements of holes in a dielectric layer, are gaining attention for their easier manufacturing via subtractive techniques, unnecessity of substrates, and access to resonant near-fields. Despite their practical relevance, their theoretical description remains elusive. Here, we present a semi-analytical dipole-quadrupole model for the multipole analysis of numerically-obtained reflection and transmission spectra in metasurfaces excited at arbitrary angles. Dipole models are generally sufficient to study traditional metasurfaces made of solid nanostructures. However, the inclusion of electric and magnetic quadrupoles is necessary to study membrane-metasurfaces, which offer an ideal platform to showcase our method. We demonstrate the importance of choosing the optimal position of a symmetric membrane-metasurface’s unit cell to ensure the sufficiency of the dipole-quadrupole approximation. We show that our formalism can explain complex phenomena arising from inter-multipole interference, including lattice anapole and generalized Kerker effects, Fano resonances, and quasi-bound states in the continuum. We also present the applicability of the method to membrane-metasurfaces with non-centrosymmetric unit cells (e.g., conical holes and surface voids). By enabling a deeper insight into the coupling mechanisms leading to the formation of local and collective resonances, our method expands the electromagnetics toolbox to study, understand, and design conventional and membrane-metasurfaces.

\end{abstract}

\section{Introduction \label{sec:intro}}

In modern optics, dielectric metasurfaces (i.e., two-dimensional arrays of resonant building blocks) offer a compact strategy to control the properties of light beyond what is possible with flat surfaces~\cite{schulz2024roadmap,babicheva2024mie}. Typical applications include ultra-thin metalenses~\cite{arbabi2023advances}, metamirrors~\cite{matiushechkina2024perfect}, polarization converters~\cite{yang2023broadband}, sensitive bio and chemical sensors~\cite{leitis2021wafer}, efficient nonlinear generation and control~\cite{vabishchevich2023nonlinear}, photovoltaics~\cite{liu2018planar,rana2023broadband}, and nanolasers~\cite{yang2021low,droulias2022experimental}, to name a few.
Conventional metasurfaces are formed by individual nanostructures on a substrate. In some cases (e.g., for collective resonances), the dielectric contrast between the substrate and superstrate is undesired and can be reduced using an index-matching superstrate (e.g., polymers). Although a superstrate limits the physical accessibility to the region of strong near fields, its addition can also enable optical tunability based on the dynamic control of the above introduced dielectric contrast~\cite{allayarov2024dynamic}.

An emerging class of metasurfaces is based on a complementary approach: instead of solid nanostructures, a periodic array of through holes of subwavelength diameter is realized in a dielectric membrane of subwavelength thickness~\cite{yang2020mie,xu2022enhanced}. These structures, called membrane metasurfaces, do not require a supporting substrate~\cite{adi2024trapping}, thus avoiding substrate-induced unwanted energy losses and shifts in the resonant frequencies of the metasurface. The resonant concentration of energy in the holes makes them ideal devices for efficient second- and third-harmonic generation~\cite{qu2022giant,zheng2023third,konishi2020circularly,konishi2020tunable}, polarization-switching~\cite{yang2020polarization}, biochemical sensing~\cite{adi2024trapping,rosas2025enhanced}, nanoplastic detection~\cite{ludescher2025optical} and even for the shaping of extreme ultraviolet radiation~\cite{Ossiander2023}. 
Moreover, membrane metasurfaces are especially suited for all those effects that require a homogeneous environment, such as surface lattice resonances (SLRs) and quasi bound states in the continuum (quasi-BIC) resonances, including their tunability by immersing the membrane, for example, in a thermally or electro-optically tunable medium~\cite{adi2024trapping}.

In general, the surge of interest in ``negative" structures is due to the possibility of directly accessing the resonant near fields inside the holes, as well as easier fabrication procedures via subtractive techniques (e.g., etching and focused ion beam milling). This also includes air voids realized on a surface~\cite{Hentschel2023} or encapsulated within a material~\cite{Avishek2025}. Despite the benefits of ``negative" structures, the theoretical approaches for their modelling are underdeveloped. For example, Babinet's principle, which applies well to metals, is not readily available for dielectrics~\cite{Hamidi2025}.
Analytical and semi-analytical multipole methods, which are important for the theoretical study of collective and coupling effects in metasurfaces, are also unavailable for the modelling of membrane metasurfaces, whose investigation has been limited frequently to numerical simulations and analysis of near-field distributions. In fact, when the characteristic size $l$ of a dielectric nanoparticle is much smaller than the incident wavelength $\lambda$ (all wavelengths $\lambda$ through the paper are in vacuum), the system exhibits mostly electric and magnetic dipole response, and one can model the metasurface's optical response via the well-known coupled-dipole model (CDM) for normal~\cite{garcia2007colloquium,evlyukhin2010optical,chen2017general} and oblique~\cite{Abujetas2020coupled,allayarov2025analytical} incidence, respectively. However, when $l/\lambda \geq 1$ or the material occupies the entire unit cell, the CDM is not sufficient to explain all observed spectral features. This is the case of membrane metasurfaces, where higher order multipoles are supported (e.g., electric and magnetic quadrupoles, and higher)~\cite{allayarov2024multiresonant}, and their interaction must be properly considered by developing specific models beyond the CDM.

Analytical models for systems with dipole-quadrupole coupling were constructed for metasurfaces of spherical particles under the condition of normal~\cite{evlyukhin2012collective,babicheva2019analytical,allayarov2024anapole} and oblique~\cite{zhu2023multipole} incidence of light. In the case of particles of arbitrary shape, one has to resort to hybrid semi-analytical methods~\cite{terekhov2019multipole}, in which the values of the multipole moments of particles in metasurfaces are first calculated numerically and then substituted into analytical expressions for the reflection and transmission coefficients. Such methods have recently been used for metasurfaces composed of individual non-spherical particles under normal incidence~\cite{allayarov2024anapole,prokhorov2022resonant,allayarov2024multiresonant}. 
However, a coupled dipole-quadrupole approach generalized to oblique incidence, which is necessary to fully characterize the behaviour of membrane metasurface, is not available and is the topic of this work.

In this paper, we develop  and demonstrate a new semi-analytical  dipole-quadrupole method (DQM) to analyze resonant spectral features of metasurfaces calculated numerically for irradiation at arbitrary angles and planes of incidence.  
In contrast to the CDM, our approach allows us to investigate metasurfaces where dipole-quadrupole and quadrupole-quadrupole interactions are taken into account at any angle of incidence.
Since our method is not focused on a specific type of particles (building blocks), it is applicable to the analysis of metasurfaces composed of particles of any shape (symmetry). The main condition for applicability is the sufficiency of the dipole-quadrupole approximation. Regardless of diffraction, our method allows us to study in detail the specular reflection and transmission of the zero diffraction order. 
Here, our semi-analytical method is efficiently demonstrated on membrane metasurfaces. 
Due to the independence of our method from the shape of the metasurface building block, we demonstrate  its applicability  to membrane metasurfaces not only with uniform cross-sectional holes, such as cylindrical or rectangular, but also with non-uniform cross-sectional holes, e.g., conical. Furthermore, holes do not necessarily need to be through, but also partially perforated in the membrane. The latter systems (i.e., with conical or partially perforated holes) are bianisotropic membrane metasurfaces (BMM) and can have important applications in designing reflectarray and controllable near-fields. In the context of recent results on the use of quasi-BIC resonances of asymmetric membrane metasurfaces \cite{rosas2025enhanced} and membranes with hole super-lattices  for sensor applications \cite{adi2024trapping}, we demonstrate the possibility to excite similar quasi-BICs in simple (one hole per unit cell) symmetric  metasurfaces due to the angular incidence of an external wave.

The paper is organized as follows: in Sec.~\ref{sec:dqmodel}, we derive reflection and transmission coefficients expressed via dipole and quadrupole moments for both TE and TM polarized incidence. In Sec.~\ref{sec:unitcell}, we investigate a dependence of the multipole decomposition on the metasurface unit-cell configuration and identify an optimal unit-cell form. In Sec.~\ref{sec:examples}, we demonstrate the accuracy of our dipole-quadrupole approach by considering a membrane metasurface consisting of cylindrical holes and excited under oblique incidence. In Sec.~\ref{sec:cases}, we discuss and analyze several interesting resonance features such as ``anti-Fano'' and angle-independent resonances of membrane metasurfaces. Additionally, we demonstrate the applicability of our approach to BMMs and metasurfaces with double-elliptical holes per unit cell. Finally, conclusions are given in Sec.~\ref{sec:concl}.

\section{Analytical model of  reflectance and transmittance \label{sec:dqmodel}}

In this section, we derive expressions for the reflection and transmission coefficients within the dipole-quadrupole approach for both TE and TM polarized incidence at an arbitrary incident angle. This is a fundamental extension of the pure dipole model~\cite{allayarov2025analytical} by adding electric (EQ) and magnetic (MQ) quadrupole moments.

\begin{figure}[!b]
\centering
\includegraphics[width=0.5\linewidth]{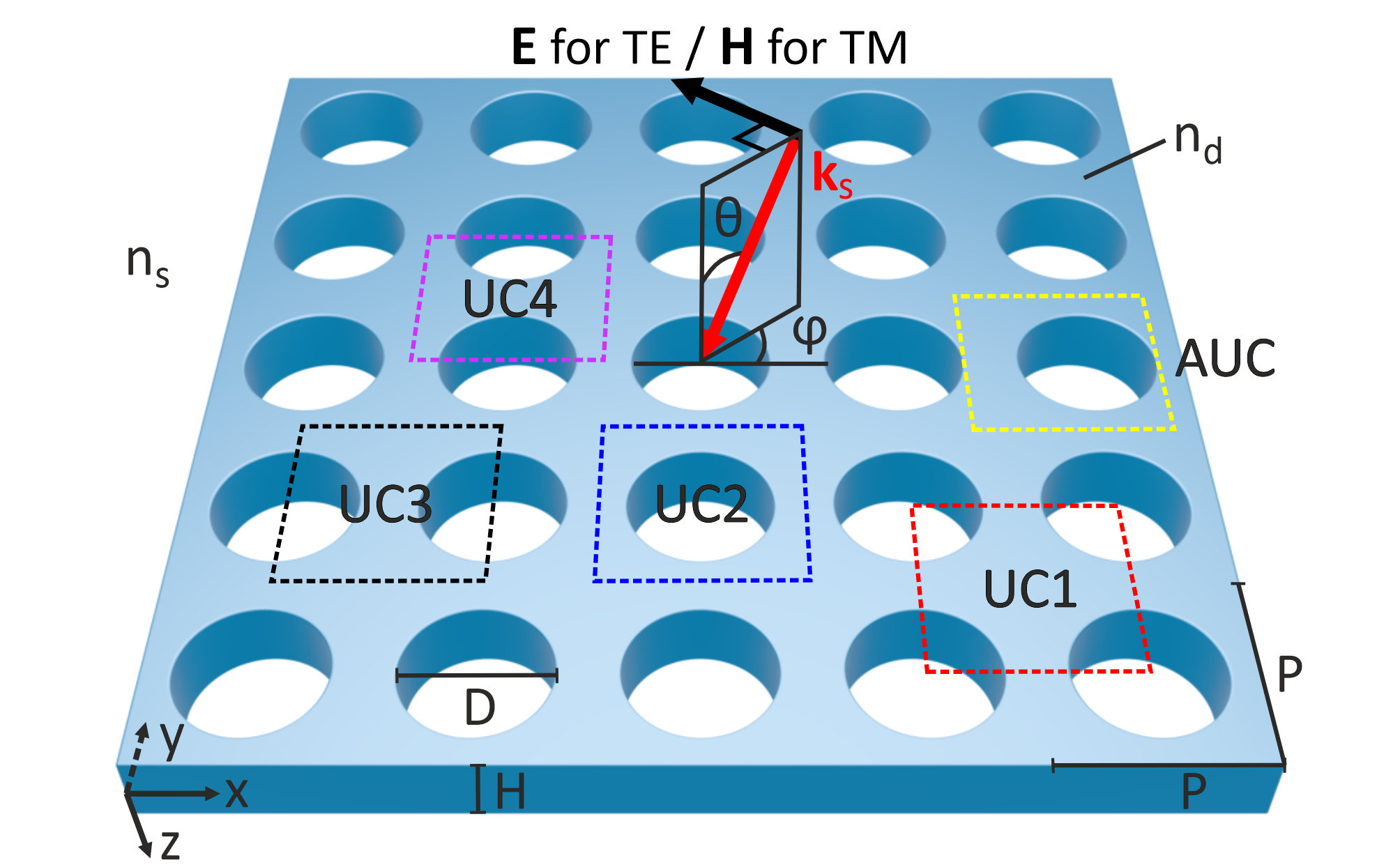}
\caption{Schematics of a membrane metasurface (alias ``holey-metasurface'') made of perforated circular (diameter $D$) periodic (period $P$) holes on dielectric layer (thickness $H$) and respective irradiation condition. The refractive index of the dielectric membrane and surrounding environment is $n_{\rm d}$ and $n_{\rm s}$, respectively. The metasurface's possible symmetric unit cell (UC) configurations are indicated by (UC1) red, (UC2) blue, (UC3) black and (UC4) purple dashed squares. The unit cell can have an asymmetric shape, e.g, the yellow dashed square illustrates one of the possible asymmetric unit cell (AUC). The arrows ($x$,$y$,$z$) indicate the direction of the Cartesian coordinate axes and the $xy$-plane is at $z=0$, i.e., the middle of the membrane layer. Note that the origin of the coordinate system has to coincide with the center of mass of the UC.} 
\label{fig:str}
\end{figure}

First, we assume that the metasurface is placed on the $xy$-plane of the coordinate system shown in Fig.~\ref{fig:str}. The incidence wave vector is 
\begin{equation}
    {\bf k}_{\rm s} = (k_x,k_y,k_z) = k_{\rm s}(\sin\theta\cos\varphi,\sin\theta\sin\varphi,\cos\theta),
\end{equation}
where $k_{\rm s} = k_0 n_{\rm s}$ is the wavenumber in the surrounding medium with refractive index $n_{\rm s}$, $k_0=c/\omega$ is the wave number in vacuum, $\omega$ is the angular frequency, $c$ is the speed of light in vacuum, $\theta$ and $\varphi$ are the polar and azimuthal angles of the incident plane wave. In our geometry (Fig.~\ref{fig:str}), the angle $\varphi$  determines the orientation of the plane of incidence, and $\theta$ is the incident angle.

We define the components of the incident plane wave's electric ${\bf E}_{\rm inc} = {\bf E}_0 \, {\rm exp}({\rm i}\mathbf{k}_{\rm s}\cdot \mathbf{r} - {\rm i}\omega t)$ and magnetic ${\bf H}_{\rm inc} = {\bf H}_0 \, {\rm exp}({\rm i}\mathbf{k}_{\rm s}\cdot \mathbf{r} - {\rm i}\omega t)$ field amplitudes as
\begin{align}\label{te1}
    &{\bf E}_0=(E_{0x},E_{0y},E_{0z})=E_0(-\sin\varphi,\cos\varphi,0), \\
    &{\bf H}_0=(H_{0x},H_{0y},H_{0z})=H_0(-\cos\theta\cos\varphi,-\cos\theta\sin\varphi,\sin\theta),
\end{align}
for TE polarization, and as
\begin{align}\label{tm1}
    &{\bf E}_0=(E_{0x},E_{0y},E_{0z})=E_0(\cos\theta\cos\varphi,\cos\theta\sin\varphi,-\sin\theta),\\
    &{\bf H}_0=(H_{0x},H_{0y},H_{0z})=H_0(-\sin\varphi,\cos\varphi,0),
\end{align}
for TM polarization. Further, for the sake of convenience, we omit the ${\rm exp}(- {\rm i}\omega t)$ factor unless it is necessary.

\subsection{TE polarization of incident waves}

Let us assume that a single MQ is located at a point with ${\bf r}_0$, if this is the coordinate system, ${\bf r}_0 = (0,0,0)$. The electric field ${\bf E}$ generated by the MQ at a coordinate ${\bf r}=(x,y,z)$ can be written as~\cite{babicheva2019analytical}:
\begin{align}
    &{\bf E}({\bf r},{\bf r}_0)=\frac{-{\rm i}k_0^2}{2\varepsilon_0 \omega}\int \hat{G}({\bf r},{\bf r}'){\bf a}({\bf r}',{\bf r}_0)d{\bf r}',\label{EMQ1}
\end{align}
where $\varepsilon_0$ is the vacuum permittivity, ${\bf a}({\bf r}',{\bf r}_0) = [\nabla \times \hat{M}\nabla\delta({\bf r}'-{\bf r}_0)]$, $\delta$ is the Dirac's delta function, $\nabla$ is the gradient operator with respect to the radius-vector ${\bf r}'$. The traceless and symmetric MQ moment tensor $\hat{M}$ has only five independent components, i.e., $M_{xx} + M_{yy} + M_{zz} = 0$, $M_{xy} = M_{yx}$, $M_{xz} = M_{zx}$ and $M_{yz} = M_{zy}$. The Green's tensor $\hat{G}({\bf r},{\bf r}')$ of a homogeneous medium with refractive index $n_{\rm s}$ can be written as~\cite{martin1998electromagnetic}:
\begin{equation}\label{GF_homo}
    \hat G({\bf r},{\bf r}')=\big(\hat U +\frac{1}{k_{\rm s}^2}\nabla\nabla \big)\phi({\bf r},{\bf r}') = \frac{1}{k_{\rm s}^2} \begin{pmatrix}
    k_{\rm s}^2+\dfrac{\partial^2}{\partial x^2} & \dfrac{\partial^2}{\partial x \partial y} & \dfrac{\partial^2}{\partial x \partial z}\\
    \dfrac{\partial^2}{\partial x \partial y} & k_{\rm s}^2+\dfrac{\partial^2}{\partial y^2} & \dfrac{\partial^2}{\partial y \partial z}\\
    \dfrac{\partial^2}{\partial x \partial z} & \dfrac{\partial^2}{\partial y \partial z} & k_{\rm s}^2+\dfrac{\partial^2}{\partial z^2}
\end{pmatrix}\phi({\bf r},{\bf r}'),
\end{equation}
where $\phi({\bf r},{\bf r}')={{\rm exp}({\rm i}k_{\rm s}|{\bf r}-{\bf r}'|})/({4\pi|{\bf r}-{\bf r}'|})$, and $\hat U$ is the $3\times 3$ unit tensor.

The components of the vector ${\bf a} = (a_x,a_y,a_z)$ are
\begin{align}\label{a1}
    &a_x({\bf r}',{\bf r}_0) = \sum_\tau \Big(M_{z\tau}\frac{\partial^2}{\partial y' \partial \tau'} - M_{y\tau}\frac{\partial^2}{\partial z' \partial \tau'} \Big)\delta({\bf r}'-{\bf r}_0),\\
    &a_y({\bf r}',{\bf r}_0) = \sum_\tau \Big(M_{x\tau}\frac{\partial^2}{\partial z' \partial \tau'} - M_{z\tau}\frac{\partial^2}{\partial x' \partial \tau'} \Big)\delta({\bf r}'-{\bf r}_0),\\
    &a_z({\bf r}',{\bf r}_0) = \sum_\tau \Big(M_{y\tau}\frac{\partial^2}{\partial x' \partial \tau'} - M_{x\tau}\frac{\partial^2}{\partial y' \partial \tau'} \Big)\delta({\bf r}'-{\bf r}_0),
\end{align}
where $\tau = (x,y,z)$, and using
\begin{align} \label{ddelta}
    &\int\delta^{(n)}(\tau'-\tau_0)f(\tau')d\tau' = (-1)^{n}f^{(n)}(\tau_0),
\end{align}
we obtain the following expression for the $x$ component of ${\bf E}$:
\begin{align}\label{Ex}
    E_x({\bf r},{\bf r}_0) =&\frac{-{\rm i}k_0^2}{2\varepsilon_0 \omega}\int \Big(G_{xx}({\bf r},{\bf r}')a_x({\bf r}',{\bf r}_0) + G_{xy}({\bf r},{\bf r}')a_y({\bf r}',{\bf r}_0) + G_{xz}({\bf r},{\bf r}')a_z({\bf r}',{\bf r}_0) \Big)d{\bf r}' \nonumber\\
     = &\frac{-{\rm i}k_0^2}{2\varepsilon_0 \omega} \sum_\tau\Big[\Big(M_{z\tau}\frac{\partial^2}{\partial y \partial \tau} - M_{y\tau}\frac{\partial^2}{\partial z \partial \tau} \Big)G_{xx}({\bf r},{\bf r}_0) \nonumber\\
    & + \Big(M_{x\tau}\frac{\partial^2}{\partial z \partial \tau} - M_{z\tau}\frac{\partial^2}{\partial x \partial \tau} \Big)G_{xy}({\bf r},{\bf r}_0) \nonumber\\
    & + \Big(M_{y\tau}\frac{\partial^2}{\partial x \partial \tau} - M_{x\tau}\frac{\partial^2}{\partial y \partial \tau} \Big)G_{xz}({\bf r},{\bf r}_0) \Big ],
\end{align}
where ${\bf r}$ is the observation point, and ${\bf r}_0$ is the position of a point MQ source. In Eq.~(\ref{Ex}), we used the relation $\phi({\bf r},{\bf r}')=\phi(|{\bf r}-{\bf r}'|)$ so that $\dfrac{\partial^2}{\partial x' \partial y'} \phi({\bf r},{\bf r}')=\dfrac{\partial^2}{\partial x \partial y} \phi({\bf r},{\bf r}')$, and this is true for other second (even) order partial derivatives.
Using the symmetry properties of the tensor $\hat{M}$, Eq.~\eqref{GF_homo}, grouping the same tensor components, and finally repeating the above derivations for the other components of the electric field, we obtain the following:
\begin{align}
    &E_x({\bf r},{\bf r}_0)  = \frac{-{\rm i}k_0^2}{2\varepsilon_0 \omega} \Big[-(M_{xx}+2M_{yy})\frac{\partial^2}{\partial y \partial z} - M_{xy}\frac{\partial^2}{\partial x \partial z} + M_{xz}\frac{\partial^2}{\partial x \partial y} + M_{yz}\Big(\frac{\partial^2}{\partial y^2}-\frac{\partial^2}{\partial z^2} \Big) \Big]\phi({\bf r},{\bf r}_0), \label{Ex10}\\
    &E_y({\bf r},{\bf r}_0)  = \frac{-{\rm i}k_0^2}{2\varepsilon_0 \omega} \Big[(2M_{xx}+M_{yy})\frac{\partial^2 }{\partial x \partial z} + M_{xy}\frac{\partial^2 }{\partial y \partial z} + M_{xz}\Big(\frac{\partial^2}{\partial z^2}-\frac{\partial^2}{\partial x^2} \Big) - M_{yz}\frac{\partial^2 }{\partial x \partial y}\Big]\phi({\bf r},{\bf r}_0), \label{Ey10}\\
    &E_z({\bf r},{\bf r}_0)  = \frac{-{\rm i}k_0^2}{2\varepsilon_0 \omega} \Big[(-M_{xx}+M_{yy})\frac{\partial^2 }{\partial x \partial y} + M_{xy} \Big(\frac{\partial^2}{\partial x^2} - \frac{\partial^2}{\partial y^2} \Big) -  M_{xz}\frac{\partial^2 }{\partial y \partial z} + M_{yz}\frac{\partial^2 }{\partial x \partial z}\Big]\phi({\bf r},{\bf r}_0). \label{Ez10}
\end{align}
Note that Eqs.~(\ref{Ex10}-\ref{Ez10}) are for the single MQ located at ${\bf r}_0$. In a periodic metasurface, we consider ${\bf r}_0$ as the origin of the coordinate system located in the unit cell with number $l=0$. In this case, the multipole moments of all the other cells (with $l\neq0$) and located in ${\bf r}_l$ will differ from the multipole moments of the central cell (with $l=0$) only by the phase factor ${\rm exp}({\rm i}\mathbf{k}^{\parallel}_{\rm s}\cdot \mathbf{r}_l^{\parallel})$, where $\mathbf{k}^{\parallel}_{\rm s}=(k_x,k_y)$ is the in-plane component of the incoming wave vector ${\bf k}_{\rm s}=({\bf k}^\parallel_{\rm s}, k_z)$, and  $\mathbf{r}^{\parallel}_l=(x_l,y_l)$. This is a consequence of  the Bloch's theorem~\cite{garcia2007colloquium} for periodic structures. Hence, the electric field components  generated by the MQ of the $l$-th unit cell in the metasurface are
\begin{align}
    &E_x({\bf r},{\bf r}_l)  \!=\! \frac{-{\rm i}k_0^2}{2\varepsilon_0 \omega} \Big[-(M_{xx}+2M_{yy})\frac{\partial^2}{\partial y \partial z} - M_{xy}\frac{\partial^2}{\partial x \partial z} + M_{xz}\frac{\partial^2}{\partial x \partial y} + M_{yz}\Big(\frac{\partial^2}{\partial y^2}-\frac{\partial^2}{\partial z^2} \Big) \Big]\phi({\bf r},{\bf r}_l){\rm e}^{{\rm i}\mathbf{k}^{\parallel}_{\rm s}\cdot \mathbf{r}_l^{\parallel}}, \label{Ex1}\\
    &E_y({\bf r},{\bf r}_l)  \!=\! \frac{-{\rm i}k_0^2}{2\varepsilon_0 \omega} \Big[(2M_{xx}+M_{yy})\frac{\partial^2 }{\partial x \partial z} + M_{xy}\frac{\partial^2 }{\partial y \partial z} + M_{xz}\Big(\frac{\partial^2}{\partial z^2}-\frac{\partial^2}{\partial x^2} \Big) - M_{yz}\frac{\partial^2 }{\partial x \partial y}\Big]\phi({\bf r},{\bf r}_l){\rm e}^{{\rm i}\mathbf{k}^{\parallel}_{\rm s}\cdot \mathbf{r}_l^{\parallel}}, \label{Ey1}\\
    &E_z({\bf r},{\bf r}_l)  \!=\! \frac{-{\rm i}k_0^2}{2\varepsilon_0 \omega} \Big[(-M_{xx}+M_{yy})\frac{\partial^2 }{\partial x \partial y} + M_{xy} \Big(\frac{\partial^2}{\partial x^2} - \frac{\partial^2}{\partial y^2} \Big) -  M_{xz}\frac{\partial^2 }{\partial y \partial z} + M_{yz}\frac{\partial^2 }{\partial x \partial z}\Big]\phi({\bf r},{\bf r}_l) {\rm e}^{{\rm i}\mathbf{k}^{\parallel}_{\rm s}\cdot \mathbf{r}_l^{\parallel}}. \label{Ez1}
\end{align}
The total field at a point $\bf r$ outside the metasurface, created by all MQs located at the lattice nodes, can be formally written as the following superposition 
\begin{equation}\label{Sup}
    {\bf E}^{\rm MQ}({\bf r}) = \sum\limits_{l=0}^\infty {\bf E}({\bf r},{\bf r}_l) \equiv \sum\limits_{l=0}^\infty {\bf E}_l({\bf r}).
\end{equation}
Thus, the above presentation of the electric field, Eq.~(\ref{Sup}), allows us to write the real space derivatives of $\phi$ in the inverse space by using the Weyl representation of a spherical wave~\cite{weyl1919ausbreitung,mandel1995optical,evlyukhin2010optical}:
\begin{align}
&\sum_{l=0}^\infty \frac{\partial^{(u+v+w)}}{\partial x^u \partial y^v \partial z^w}\phi({\bf r},{\bf r}_l){\rm e}^{{\rm i}\mathbf{k}^{\parallel}_{\rm s}\cdot \mathbf{r}_l^{\parallel}}=\frac{{\rm i}}{2S_{\rm L}}\sum^\infty_{\bf L} (-{\rm i}\tilde{L}_x)^u(-{\rm i}\tilde{L}_y)^v(\pm{\rm i}\tilde{L}_z)^w \tilde{F},\label{dphi1}
\end{align}
 where ``$-$" corresponds to $z<0$ (reflection), ``$+$" corresponds to $z>0$ (transmission),  
\begin{align}
&\tilde{F} = \frac{1}{\tilde{L}_z}{\rm e}^{{\rm i}(-\tilde{L}_xx-\tilde{L}_yy\pm\tilde{L}_zz)},\\
&\tilde{L}_x = L_x - k_x = L_x - k_{\rm s}\sin\theta\cos\varphi,\\
&\tilde{L}_y = L_y - k_y = L_y - k_{\rm s}\sin\theta\sin\varphi,\\
&\tilde{L}_z = \sqrt{k_{\rm s}^2-\tilde{L}_x^2-\tilde{L}_y^2},
\end{align}
${\bf L} = (L_x,L_y,L_z)$ is the lattice vector in the inverse space, and $S_{\rm L}$ is the unit cell area {of the metasurface} in the real space. In Eqs.~(\ref{Sup}) and (\ref{dphi1}), for the sake of our convenience, we have numbered the lattice nodes (unit cells) $l=0,1,2,...$ and assumed that the node $l=0$ is at $\mathbf{r}_0=0$. The total electric field generated by all magnetic quadrupoles is the superposition of the field of each magnetic quadrupole located at every lattice node. Therefore, the sum starts from $l=0$. Since we consider an infinite periodic array, the upper limit of the sum is infinity. We use this notation throughout the work. Furthermore, full numerical simulations with periodic boundary conditions correspond to the sum from $l=0$ to $l=\infty$.

For the specular {(zero order diffraction)} reflection and transmission, i.e., ${\bf L} = 0$:
\begin{align}
&\sum_{l=0}^\infty \frac{\partial^{(u+v+w)}}{\partial x^u \partial y^v \partial z^w}\phi({\bf r},{\bf r}_l){\rm e}^{{\rm i}\mathbf{k}^{\parallel}_{\rm s}\cdot \mathbf{r}_l^{\parallel}} \approx \frac{{\rm i}}{2S_{\rm L}} \frac{({\rm i}k_{\rm s}\sin\theta\cos\varphi)^u({\rm i}k_{\rm s}\sin\theta\sin\varphi)^v(\pm{\rm i}k_{\rm s}\cos\theta)^w}{k_{\rm s}\cos\theta}{\rm e}^{{\rm i}(k_xx+k_yy \pm k_zz)}.\label{dphi1L0}
\end{align}
Note that for radiation with a wavelength exceeding the wavelength of the first-order Rayleigh anomaly of the metasurface, the zeroth-order diffraction provides the only contribution to the far-field zone.

By substituting Eq.~(\ref{dphi1L0}) into {Eq.~(\ref{Sup})}, we obtain the following expressions for the electric field:
\begin{align} 
    E_{\parallel}^{\rm MQ} = &\frac{{\rm i}k_{\rm s}\:{{\rm e}^{{\rm i}(k_xx+k_yy \pm k_zz)}}}{2S_{\rm L}\varepsilon_0\varepsilon_{\rm s}}\Big(\frac{-{\rm i}k_{\rm s}}{2v} \Big) \Big[\Big( -\frac{1}{2}(M_{xx}-M_{yy})\sin2\varphi + M_{xy}\cos2\varphi \Big)(\pm \sin\theta) \nonumber\\
    & + (-M_{xz}\sin\varphi  + M_{yz}\cos\varphi )\cos\theta \Big], \label{EMQpll} \\
    E_{\perp}^{\rm MQ} = &\frac{{\rm i}k_{\rm s}\:{{\rm e}^{{\rm i}(k_xx+k_yy \pm k_zz)}}}{2S_{\rm L}\varepsilon_0\varepsilon_{\rm s}\cos\theta}\Big(\frac{-{\rm i}k_{\rm s}}{2v} \Big)\Big[-\frac{1}{2}\Big(\frac{3}{2}(M_{xx}+M_{yy})+\frac{1}{2}(M_{xx}-M_{yy})\cos2\varphi \nonumber\\ & + M_{xy}\sin2\varphi \Big)(\pm \sin2\theta) - \Big(M_{xz}\cos\varphi + M_{yz}\sin\varphi \Big)\cos2\theta \Big], \label{EMQper} \\
    E_{z}^{\rm MQ} = &\frac{{\rm i}k_{\rm s}\:{{\rm e}^{{\rm i}(k_xx+k_yy \pm k_zz)}}}{2S_{\rm L}\varepsilon_0\varepsilon_{\rm s}\cos\theta}\Big(\frac{-{\rm i}k_{\rm s}}{2v} \Big)\Big[-\Big(-\frac{1}{2}(M_{xx}-M_{yy})\sin2\varphi + M_{xy} \cos2\varphi\Big)\sin^2\theta \nonumber \\
    &-(\pm) \Big(-M_{xz}\sin\varphi + M_{yz}\cos\varphi \Big)\sin\theta\cos\theta \Big ], \label{EMQZ}
\end{align}
where ``$-$" sign corresponds to $z<0$, ``$+$" sign corresponds to $z>0$.

In the above equations $v=c/n_{\rm s}$ is the speed of light in the {surrounding} medium with refractive index $n_{\rm s}$, and for the sake of convenience, we have introduced the parallel $E_{\parallel}$ and perpendicular $E_{\perp}$ components of the electric field with respect to the plane of incidence, which are defined {through the field components $E_x$ and $E_y$ in the coordinate system shown in Fig.\ref{fig:str}} as 
\begin{equation}
\begin{pmatrix}
    E_\parallel \\
    E_\perp
\end{pmatrix}
=
R_{\rm m}
\begin{pmatrix}
    E_x  \\
    E_y
\end{pmatrix}, 
\quad 
R_{\rm m} = \begin{pmatrix}
    \cos\varphi & \sin\varphi\\
    -\sin\varphi & \cos\varphi
\end{pmatrix},
\end{equation}
where $R_{\rm m}$ is the rotation matrix which rotates the plane of incidence counterclockwise in the $xy$-plane by the angle $\varphi$. In Eqs.~(\ref{EMQpll}-\ref{EMQZ}), one has to use ``$-$" for the reflected (when $z<0$) field and ``$+$"  for the transmitted (when $z>0$) field.

Thus, repeating the same derivations for the electric field of EQ (see Appendix~\ref{sec:appA}) and using the electric fields of electric dipoles (ED) and magnetic dipoles (MD) obtained in Ref.~\cite{allayarov2025analytical}, we have the total electric field generated by {dipoles and quadrupoles moments at a point ${\bf r}$ outside the metasurface:
\begin{equation}
    {\bf E}^{\rm Total}({\bf r})={\bf E}^{\rm ED}({\bf r})+{\bf E}^{\rm MD}({\bf r})+{\bf E}^{\rm EQ}({\bf r})+{\bf E}^{\rm MQ}({\bf r}).
\end{equation}}
Specular reflection coefficients of the electric field can be obtained from the following definition:
\begin{equation}
    r^{\rm TE}_\perp=\frac{E_\perp^{\rm Total}}{E_{0\perp}{\rm e}^{{\rm i}{\bf k}_{\rm s}\cdot{\bf r}}}{\Big|_{z=0}}, \quad 
    r^{\rm TE}_\parallel=\frac{E_\parallel^{\rm Total}}{E_{0\perp}{\rm e}^{{\rm i}{\bf k}_{\rm s}\cdot{\bf r}}}{\Big|_{z=0}}, 
\end{equation}
where $E^{\rm Total}_{\parallel}$ and $E^{\rm Total}_{\perp}$ are the parallel and perpendicular components of the total reflected electric field for $z<0$, respectively; $E_{0\perp} = E_{0y}\cos\varphi - E_{0x}\sin\varphi = E_0$ is the amplitude of the electric field   of the incident wave, so that $E_{0\parallel} =  0$.

Finally, we obtain
\begin{align}
  r^{\rm TE}_\perp (\theta,\varphi) = &\frac{{\rm i}k_{\rm s}}{2S_{\rm L}{E}_0\varepsilon_0\varepsilon_{\rm s}\cos\theta}\Big[p^\perp_2+\frac{1}{v}m^\parallel_1\cos\theta + \frac{1}{v}m_3\sin\theta  \nonumber\\
  &+\frac{{\rm i} k_{\rm s}}{6}\Big(Q_3^\perp\cos\theta - Q_1^\perp\sin\theta\Big) +\frac{{\rm i} k_{\rm s}}{2v}\Big(M_3^\parallel\cos 2\theta - \tilde{M}_1^\parallel\sin2\theta\Big)\Big], \label{rDQ_TE_perp} \\
  r^{\rm TE}_\parallel (\theta,\varphi) = &\frac{{\rm i}k_{\rm s}}{2S_{\rm L}{E}_0\varepsilon_0\varepsilon_{\rm s}}\Big[p^\parallel_1\cos\theta+p_3\sin\theta-\frac{1}{v}m^\perp_2 \nonumber\\
  &+\frac{{\rm i} k_{\rm s}}{6}\Big(Q_3^\parallel\cos2\theta - \tilde{Q}_1^\parallel\sin2\theta\Big) +\frac{{\rm i} k_{\rm s}}{2v}\Big(-M_3^\perp\cos \theta + M_1^\perp\sin\theta\Big)\Big]. \label{rDQ_TE_pll}
\end{align}
In the above equations, ${\bf p} = (p_x,p_y,p_z)$ and ${\bf m} = (m_x,m_y,m_z)$ are the electric and magnetic dipole moments, $\hat{Q}$ is the traceless and symmetric electric quadrupole moment tensor, $p_3=p_z$, $m_3=m_z$, and 
\begin{equation}\label{pmD}
\begin{pmatrix}
    p^\parallel_1 \\
    p^\perp_2
\end{pmatrix}
=
R_{\rm m}
\begin{pmatrix}
    p_x  \\
    p_y
\end{pmatrix},
\quad 
\begin{pmatrix}
    m^\parallel_1 \\
    m^\perp_2
\end{pmatrix}
=
R_{\rm m}
\begin{pmatrix}
    m_x  \\
    m_y
\end{pmatrix},
\end{equation}

\begin{equation}\label{EH2}
\begin{pmatrix}
    Q^\parallel_3 \\
    Q^\perp_3
\end{pmatrix}
=
R_{\rm m}
\begin{pmatrix}
    Q_{xz}  \\
    Q_{yz}
\end{pmatrix},
\quad 
\begin{pmatrix}
    M^\parallel_3 \\
    M^\perp_3
\end{pmatrix}
=
R_{\rm m}
\begin{pmatrix}
    M_{xz}  \\
    M_{yz}
\end{pmatrix},
\end{equation}

\begin{equation}
\begin{pmatrix}
    Q^\parallel_1 \\
    Q^\perp_1
\end{pmatrix}
=
R_{\rm m}^2
\begin{pmatrix}
    \dfrac{1}{2}[Q_{xx}-Q_{yy}]  \\
    Q_{xy}
\end{pmatrix},
\quad
\begin{pmatrix}
    M^\parallel_1 \\
    M^\perp_1
\end{pmatrix}
=
R_{\rm m}^2
\begin{pmatrix}
    \dfrac{1}{2}[M_{xx}-M_{yy}]  \\
    M_{xy}
\end{pmatrix},
\end{equation}

\begin{align}
    \tilde{Q}_1^\parallel = \frac{1}{2}\Big[\frac{3}{2}(Q_{xx}+Q_{yy}) + Q_1^\parallel \Big],
\quad
  \tilde{M}_1^\parallel = \frac{1}{2}\Big[\frac{3}{2}(M_{xx}+M_{yy}) + M_1^\parallel \Big].
\end{align}
The subscripts (1,2) and (3) denote the in-plane ($x$,$y$) and out-of-plane ($z$) components, respectively. An explicit form of the transmission coefficients is presented in Appendix~\ref{sec:appB}.

{\subsection{TM polarization of incident waves}}

It is more convenient to calculate the reflection coefficients for TM-polarized incidence via magnetic fields:
\begin{equation}
    r^{\rm TM}_\perp=\frac{H_\perp^{\rm Total}}{H_{0\perp}{\rm e}^{{\rm i}{\bf k}_{\rm s}\cdot{\bf r}}}{\Big|_{z=0}}, \quad 
    r^{\rm TM}_\parallel=\frac{H_\parallel^{\rm Total}}{H_{0\perp}{\rm e}^{{\rm i}{\bf k}_{\rm s}\cdot{\bf r}}}{\Big|_{z=0}}, 
\end{equation}
where $H^{\rm Total}_{\parallel}$ and $H^{\rm Total}_{\perp}$ are the parallel and perpendicular components of the total reflected magnetic field {for $z<0$}, respectively, and they can be calculated by using ${\bf H}^{\rm Total} = \dfrac{c\varepsilon_0}{{\rm i}k_0} \nabla \times {\bf E}^{\rm Total}$. 

It is interesting to note that the reflection coefficients for TM and TE polarizations are related to each other via: 
\begin{align} \label{rDQ_TM}
  &r^{\rm TM}_\perp = -\frac{r^{\rm TE}_\parallel}{\cos\theta},\quad r^{\rm TM}_\parallel = r^{\rm TE}_\perp\cos\theta.
\end{align}
Corresponding relations for the transmission coefficients can be found in Appendix~\ref{sec:appB}. The reflectance is $R = |r_\perp|^2 + |r_\parallel|^2$.

\subsection{Special cases}

In the case of $\varphi=0$, when the plane of incidence coincides with the $xz$-plane (Fig.~\ref{fig:str}), the reflection coefficients can be written as:
\begin{align}\label{rTE}
  r^{\rm TE} = &\frac{{\rm i}k_{\rm s}}{2S_{\rm L}{E}_0\varepsilon_0\varepsilon_{\rm s}\cos\theta}\Big( \Big[ p_y + \frac{1}{v}m_z\sin\theta - \frac{{\rm i}k_{\rm s}}{6} Q_{xy} \sin\theta + \frac{{\rm i}k_{\rm s}}{2v} M_{xz} \cos2\theta \Big] \nonumber\\
  &+\Big[\frac{1}{v}m_x\cos\theta + \frac{{\rm i}k_{\rm s}}{6} Q_{yz} \cos\theta - \frac{{\rm i}k_{\rm s}}{2v} (M_{xx} + \frac{1}{2} M_{yy})\sin2\theta \Big] \Big),
\end{align}
\begin{align}\label{rTM}
  r^{\rm TM} = &\frac{{\rm i}k_{\rm s}}{2S_{\rm L}{E}_0\varepsilon_0\varepsilon_{\rm s}\cos\theta}\Big( \Big[ \frac{1}{v}m_y - p_z\sin\theta - \frac{{\rm i}k_{\rm s}}{6} Q_{xz} \cos2\theta - \frac{{\rm i}k_{\rm s}}{2v} M_{xy} \sin\theta \Big] \nonumber\\
  &+\Big[-p_x\cos\theta + \frac{{\rm i}k_{\rm s}}{6} (Q_{xx} + \frac{1}{2} Q_{yy}) \sin2\theta + \frac{{\rm i}k_{\rm s}}{2v} M_{yz} \cos\theta \Big] \Big).
\end{align}
The above Eqs.~\eqref{rTE} and \eqref{rTM} include only excited (nonzero) dipole and quadrupole moments, and the coupled moments are grouped into the square brackets. {The zeros of the expressions inside the square brackets at certain angle $\theta$ (including $\theta=0$) and wavelength $\lambda$ correspond to  accidental BIC} \cite{allayarov2024anapole,han2024observation}. Furthermore, certain quadrupole contributions have $\sin2\theta$ or $\cos2\theta$ angular dependence. Hence, at $\theta=45^{\circ}$, the corresponding contributions can either have maximum angular contribution or vanish. For example, in Eq.~\eqref{rTE}, the term proportional to $M_{xz}$ vanishes since $\cos (2 \cdot 45^{\circ})=0$, although $M_{xz} \neq 0$. These angular dependencies can additionally be used for tuning spectral features of metasurfaces.

Note that information on which multipole moments are coupled with each other in the metasurface follows from the application of the symmetry approach given in Ref.~\cite{allayarov2024multiresonant}, taking into account oblique illumination. Specifically, in Ref.~\cite{allayarov2024multiresonant}, it is explained how and which multipoles couple under normal incidence excitation, while from Ref.~\cite{allayarov2025analytical} one can understand it for dipoles excited under oblique incidence. In the general case where high order multipoles are considered, by visualizing the spectral dependency of
multipole moments, one can identify which multipoles are coupled, since around a resonance coupled multipoles will also have a resonant response. Based on these information one can construct the Table~\ref{tab1}, which shows it for dipoles and quadrupoles in the case of arbitrary angle of incidence. 

\renewcommand{\arraystretch}{1.25}
\begin{table}[!htb]
\caption{List of coupled moments. There is no coupling between the multipole moments of subgroups 1) and 2) in the case of $\varphi=0$, $\theta \neq 0$.} 
\label{tab1}
\begin{center}
\begin{tabular}{ |l|l|l| }
\hline
 Case: $\varphi=0$, $\theta \neq 0$ & TE & TM \\ \hline
\multirow{2}{*}{Coupled ($\Leftrightarrow$) moments}
 & {1) $p_y \Leftrightarrow m_z \Leftrightarrow Q_{xy} \Leftrightarrow M_{xz}$} & {1) $m_y \Leftrightarrow p_z \Leftrightarrow M_{xy} \Leftrightarrow Q_{xz}$} \\
 & {2) $m_x \Leftrightarrow Q_{yz} \Leftrightarrow M_{xx} \Leftrightarrow M_{yy}$} & {2) $p_x \Leftrightarrow M_{yz} \Leftrightarrow Q_{xx} \Leftrightarrow Q_{yy}$} \\ 
 \hline
\multirow{4}{*}{Not excited (zero) moments}
 & {$p_x = p_z = 0$} & {$p_y = 0$} \\
 & {$m_y = 0$} & {$m_x = m_z = 0$} \\
 & {$Q_{xx} = Q_{yy} = Q_{xz} = 0$} & {$Q_{xy} = Q_{yz} = 0$} \\
 & {$M_{xy} = M_{yz} = 0$} &  {$M_{xx} = M_{yy} = M_{xz} = 0$} \\ 
 \hline
 Case: $\varphi\in(0, 90^{\circ})$, $\theta \neq 0$ & TE & TM \\ \hline
 \multirow{1}{*}{Multipole moments}
 & {All nonzero and coupled} & {All nonzero and coupled} \\
 \hline
\end{tabular}
\end{center}
\end{table}

\subsection{Procedure for practical application}

It is important to note that the expressions for the reflection and transmission coefficients obtained in this section include the dipole and quadrupole moments of only the central cell of the metasurface. In this case, the calculation of the multipole moments of this cell can be performed using the integral definitions from Appendix~\ref{sec:appC} and the distribution of the total electric field in the cell obtained by numerical methods. The obtained analytical expressions allow one to perform a multipole analysis of the reflection and transmission spectra, which in the general case can only be calculated numerically. In this general practical case, the electric field $E_0$ in the expressions of reflection and transmission coefficients (for example Eqs.~(\ref{rDQ_TE_perp}) and (\ref{rDQ_TE_pll}) and all corresponding equations below) is the electric field of the incident wave at the localization point of the dipole and quadrupole moments of the central cell of the metasurface. In our case, this value coincides with the electric amplitude of the incident wave, since we consider  ${\bf E}({\bf r})={\bf E}_0\exp{({\rm i}{\bf k}_{\rm s} {\bf r})}$, and the multipoles are localized at the origin of the coordinate system for which ${\bf r}=0$.
Thus, several basic steps in using the resulting analytical model can be identified:

\noindent{1) Using numerical methods and appropriate periodic boundary conditions, the transmission and reflection coefficients of a given metasurface are calculated for a fixed frequency of the incident wave.}

\noindent {2) The total electric field distribution in the central unit cell of the metasurface is exported from the numerical calculation.}

\noindent {3) Given the total electric field distribution and the permittivity distribution in the unit cell, we obtain the distribution of induced polarization (or current density) in the unit cell.}

\noindent {4) Using integral expressions for the dipole and quadrupole moments, their Cartesian components are calculated.}

\noindent {5) The resulting dipole and quadrupole moments and angular irradiation characteristics are substituted into analytical expressions for the reflection and transmission coefficients.}

\noindent {6) Repeating the above procedure for frequencies (wavelengths) from the selected spectral region, we obtain fully numerical and dipole-quadrupole representations of the reflectance and transmittance spectra. Comparison of these representations and analysis of the dipole-quadrupole model provides information on the mechanism of resonant features formation in the system.} We note that, in general, it is not necessary to calculate reflectance and transmittance numerically for each case and compare them with the dipole-quadrupole representations. As soon as we know that the system exhibits mostly quadrupole response (e.g., from the first comparison of full numerical and dipole-quadrupole approach or from the wavelength of interest and parameters of the structure), one can use the dipole-quadrupole approach for the multipole analysis of the spectral features of the system.

In the next section, we will show that the developed approach can be successfully applied to study the properties of membrane metasurfaces.

\section{Optimal choice of the unit cell\label{sec:unitcell}}

It is known that the multipole expansion  depends on the position of the expansion center ~\cite{raab2005multipole,ospanova2023modified}. The choice of this center is determined by the requirement of minimizing the number of multipole moments that allow one to describe the electromagnetic fields emitted by the system with the required accuracy. In the case of single particles (scatterers) with simple symmetry (cubes, disks, cones, pyramids, etc.), their center of mass is usually suitable for this. If the geometry or heterogeneity of the system becomes more complex, then choosing the optimal point of localization of multipole moments becomes a non-trivial task. As was recently shown in~\cite{kildishev2025art}, in this case, to minimize the number of multipole moments, it may be more appropriate to choose different localization points for the multipole moments of the electric and magnetic types. However, we note that this may  lead to a complication of the multipole analysis of electromagnetic fields generated (emitted) by the entire system.

Using the multipole method to analyze the properties of a metasurface (a two-dimensional periodic structure) comes down to replacing it with a two-dimensional system of point multipole moments located with the same periodicity. However, this also raises the question of choosing the point of localization of multipole moments inside the unit cell of the metasurface. In the case of metasurfaces consisting of identical particles with high symmetry, the choice of the unit cell and the point of localization of multipoles in it is quite obvious: the unit cell is chosen so that its center coincides with the center of mass (or the center of symmetry) of the particle, and this point is also chosen to localize the multipole moments. This approach corresponds to the case of a single particle, and is often used in the multipole analysis of metasurfaces made of separate particles~\cite{terekhov2019multipole,babicheva2021multipole}. This is justified by the fact that with such a choice, the particles are located in the center of the elementary cells, and therefore the values of their high-order multipole moments are sharply reduced. As a result, this leads to a minimization of the number of significant multipole moments and a simplification of the multipole analysis.

In the case of membrane metasurfaces, the situation changes. In this case, the choice of the unit cell (UC) of the metasurface requires special considerations. Indeed, unlike metasurfaces of individual particles, in the case of membranes it is impossible to choose a unit cell for which all the matter would be concentrated only in its central part. With this in mind, we examine this issue in more detail below.

\subsection{Symmetric unit cell\label{sec:unitcell1}}

As one can see from the insets in Fig.~\ref{fig:str}, it is possible to choose, for example, four different UC configurations, i.e., UC1, UC2, UC3 and UC4. It is possible to consider  that the center of mass of every UC in Fig.~\ref{fig:str} is at the origin of the coordinate system, ${\bf r}_{\rm UC} = (0,0,0)$, where its multipole moments are located. Since the distribution of matter in the cell is different for each case, the multipole decompositions must also differ from each other. Let us, for demonstration, calculate the reflection coefficient and its multipole contributions. As an example, we consider a membrane metasurface (Fig.~\ref{fig:str}) with $D=200$~nm, $H=200$~nm, $P=300$~nm, $n_{\rm d}=2.45$ (e.g., diamond or TiO$_2$ in visible spectrum), $n_{\rm s}=1$, and normal incidence ($\theta=\varphi=0$) plane wave excitation. In the condition of normal incidence, the reflection coefficient $r$ can be decomposed up to the 16-th poles~\cite{allayarov2024multiresonant}:
\begin{align} \label{r16}
r=&\frac{{\rm i}k_{\rm s}}{2S_{\rm L}{E}_0\varepsilon_0\varepsilon_{\rm s}} \Bigl(p_x - \frac{1}{v}m_y + \frac{{\rm i}k_{\rm s}}{6}Q_{xz} - \frac{{\rm i}k_{\rm s}}{2v}M_{yz} -\frac{k^2_{\rm s}}{6}O^{{\rm (e)}}_{xzz} + \frac{k^2_{\rm s}}{6v}O^{{\rm (m)}}_{yzz} - \frac{{\rm i}k^3_{\rm s}}{24}S^{{\rm (e)}}_{xzzz} + \frac{{\rm i}k^3_{\rm s}}{24v}S^{{\rm (m)}}_{yzzz} \Bigr),
\end{align}
where $\hat{O}^{\rm (e)}$, $\hat{O}^{\rm (m)}$, $\hat{S}^{\rm (e)}$ and $\hat{S}^{\rm (m)}$ are the electric octupole (EO), magnetic octupole (MO), electric 16-th pole (E16) and magnetic 16-th pole  moments of a central unit cell, respectively. Here, we would like to note that Eq.~\eqref{r16} was first obtained and applied for metasurfaces consisting of isolated (non-touching) particles~\cite{allayarov2024multiresonant}. Below, we show that the above equation and the equations obtained in this work are also applicable to the multipole representation of the reflection and transmission coefficients of dielectric membrane metasurfaces. To perform a multipole analysis, it is convenient to write the reflection coefficients as the sum of individual multipole contributions and compare each individual contribution. For example, Eq.~\eqref{r16} can be rewritten in such representation 
\begin{equation}\label{rrr}
r=r^{p_x}+r^{m_y}+r^{Q_{xz}}+r^{M_{yz}}+r^{O^{(e)}_{xzz}}+r^{O^{(m)}_{yzz}}+r^{S^{(e)}_{yzzz}} +r^{S^{(m)}_{yzzz}}\:,
\end{equation}
where each term on the right side of the equality corresponds to the corresponding  multipole. Below, when we consider multipole contributions, we will mean the absolute value of each multipole term in the reflection coefficient.

\begin{figure*}[!htb]
\centering
\includegraphics[width=\linewidth]{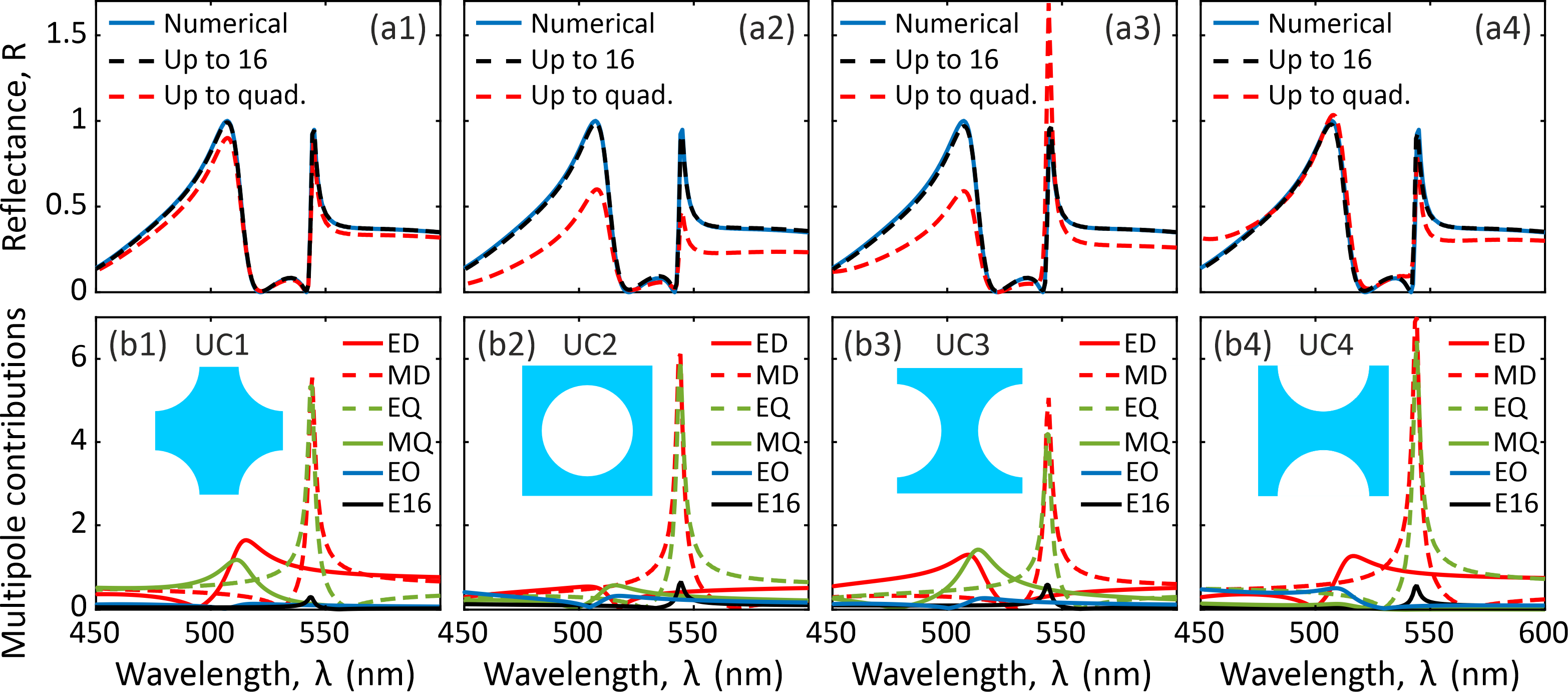}
\caption{(a) Reflectance and (b) its absolute multipole contributions of the membrane metasurface shown in Fig.~\ref{fig:str} with $H=200$~nm, $D=200$~nm, $P=300$~nm and the normal incidence excitation ($\theta=\varphi=0$). The inset in (b1-b4) shows the considered unit cell configuration, UC1-UC4, respectively. The multipole contributions are plotted in a unitless form as the absolute values of corresponding terms in  Eq.~\eqref{rrr}. The refractive index of the dielectric layer and surrounding environment is $n_{\rm d}=2.45$ and $n_{\rm s}=1$, respectively.}
\label{fig:rmuc}
\end{figure*}

In Fig.~\ref{fig:rmuc}, reflectance $R=|r|^2$ (top panels) and its main multipole contributions (bottom panels) for four different UC configurations are shown. Obviously, full numerical reflectance (solid blue curve) is the same for all cases, since it does not depend on the shape of the unit cell. The reflectance based on Eq.~\ref{r16} (dashed black curve) has excellent agreement with the full numerical one in all cases. Interestingly, if we truncate the decomposition in Eq.~\ref{r16} up to the quadrupoles (which is equivalent to the case of this paper) and compare the reflectance (dashed red curves) with the previous two, a very different behavior is observed depending on the UC configuration. The reflectance of the quadrupole approximation based on UC1 is the closest to the exact one. This means that in the case of UC1 [see Fig.~\ref{fig:rmuc}(b1)], multipoles other than dipoles and quadrupoles, i.e., octupoles and 16-th poles have a negligible contribution. Second, in the case of UC2-UC4 [see Fig.~\ref{fig:rmuc}(b2)-(b4)], the multipole contributions and content are significantly different from the UC1 case. For example, in the case of UC2, around $\lambda=510$~nm, ED and MQ resonances are almost suppressed, and the E16 resonance at $\lambda=545$~nm is comparable with them. Furthermore, the Fano profile of the multipole resonances flips over, i.e., the zeros of the multipole contributions (anapole states) occur on the other side of the resonance position. Thus, UC1 is the most optimal UC configuration among the considered ones and provides a faster convergence of reflection in terms of number of multipoles. 
Therefore, in our calculation below, we will use UC1. 

The explanation of why the choice of the UC1 cell gives the best agreement of the dipole-quadrupole model with the full numerical calculation follows from the distribution of the matter in the cell. It is evident from the inset in Fig.~\ref{fig:rmuc}(b1) that a greater portion of the UC's material is concentrated in the central part of the cell. It follows that the multipole moments are determined mainly by the induced currents in the central part of the cell. Since this region is small compared to the full size of the cell and the wavelength of the radiation, the response of the system is determined only by the dipoles and quadrupoles. For other types of cells UC2, UC3, UC4 the matter is concentrated on the edge regions of the cell, which leads to a significant increase in the role of the multipole moments of higher orders.

\subsection{Asymmetric unit cell\label{sec:unitcell2}}

\begin{figure*}[t]
\centering
\includegraphics[width=0.85\linewidth]{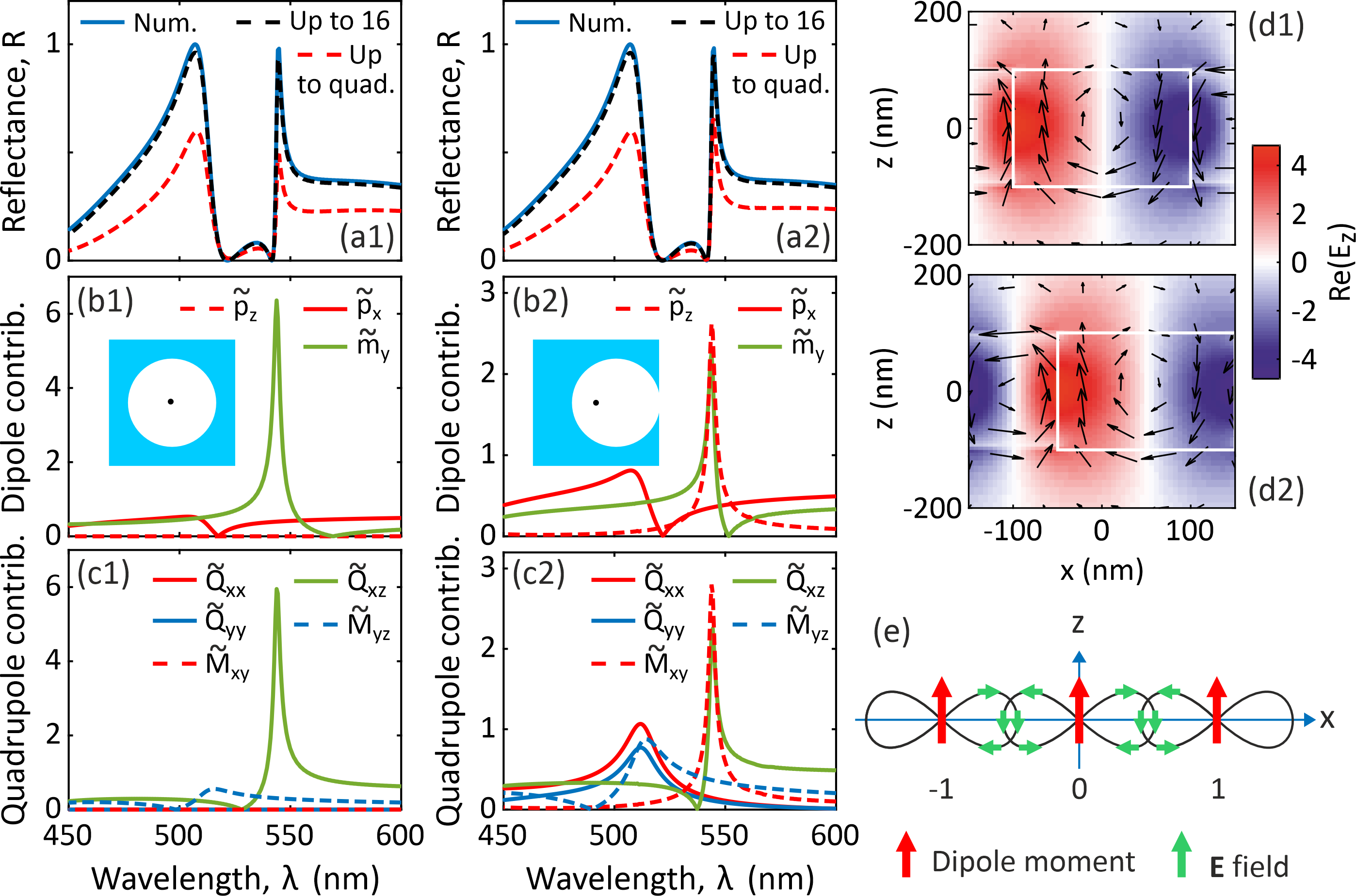}
\caption{(a) Reflectance and its (b) {absolute} dipole and (c) {absolute} quadrupole contributions of the membrane metasurface under the conditions considered in Fig.~\ref{fig:rmuc}. The inset in (b) shows the considered unit cell configuration: (1) symmetric and (2) asymmetric unit cell. The black dot indicates the center of multipole decomposition. In (d1) and (d2), it is shown the distribution of the real part of the $z$-component of the electric field within symmetric and asymmetric unit cells ($xz$-plane, $y=0$), respectively, at the wavelength $\lambda=545$~nm. The black arrows indicate the direction of the electric field. (e) Schematic presentation of the scattering of an array of out-of-plane dipole moments in the $xz$-plane. }
\label{fig:m_bian}
\end{figure*}

There is another important reason why an optimal choice of a unit cell is crucial to properly explain the spectral features of a metasurface. We note that all considered unit cells in Fig.~\ref{fig:rmuc} are symmetrical. 
In that case, only certain components of multipole moments (within the independent components) are nonzero. We call them the contributing moments. 
They directly contribute to the reflection and transmission coefficients. Indeed, as shown in Fig.~\ref{fig:m_bian}(a1)-(c1) for the symmetrical UC2 within the quadrupole approximation, only in-plane dipole moments $p_x$ and $m_y$, and ``out-of-plane'' quadrupole moments $Q_{xz}$ and $M_{yz}$ are nonzero. 
On the other hand, the unit cell can be asymmetric. The panels (a2-c2) of Fig.~\ref{fig:m_bian} depict the previous results for the unit cell in which the hole is shifted to $R/2=50$~nm along the $x$-axis [see inset of Fig.~\ref{fig:m_bian}(b2)]. 
As expected, the reflectance does not change, despite the fact that the amplitudes of the contributing moments are approximately twice smaller than before. However, additional out-of-plane electric dipole moment $p_z$ and ``in-plane'' quadrupole moments $Q_{xx}$, $Q_{yy}$ and $M_{xy}$ are excited and very significantly. This is due to the fact that the electric field distribution in the unit cell is asymmetric with respect to the decomposition point [see Fig.~\ref{fig:m_bian}~(d1) and (d2)]. Interestingly, different components of the same multipole moment can have resonances at different wavelengths. For example, the resonance peaks of $p_x$ and $p_z$ are around $\lambda = 500$~nm and $\lambda = 550$~nm, respectively. To understand such multipole moment redistribution, as an example, let us consider the scattered electric field by a single electric dipole and magnetic quadrupole. The scattered electric far-field is proportional to the following expression~\cite{evlyukhin2016optical}:
\begin{equation}\label{Esca}
\mathbf{E} \propto [\mathbf{n} \times [\mathbf{p} \times \mathbf{n}]] + \frac{{\rm i}k_{\rm s}}{2v}[\mathbf{n} \times (\hat{M}\mathbf{n})],
\end{equation}
where $\mathbf{n}=(n_x, n_y, n_z)$ is the unit vector, which indicates the scattering direction. It is important to note that under normal excitation, multipoles located at each lattice point will have the same phase. From Fig.~\ref{fig:m_bian}(b2,c2), we see that around $\lambda = 550$~nm $p_z$ and $M_{xy}$ are in resonance. Therefore, we expand Eq.~\eqref{Esca} and keep the terms containing only these moments:
\begin{align}
E_x &\propto -\big(p_z + \frac{{\rm i}k_{\rm s}}{2v}M_{xy} \big) n_x n_z, \label{Esca2x}\\
E_y &\propto \big(-p_z + \frac{{\rm i}k_{\rm s}}{2v}M_{xy} \big) n_y n_z, \label{Esca2y}\\
E_z &\propto \big(p_z + \frac{{\rm i}k_{\rm s}}{2v}M_{xy} \big) n_x^2 + \big(p_z - \frac{{\rm i}k_{\rm s}}{2v}M_{xy} \big) n_y^2. \label{Esca2z}
\end{align}
We observe that $p_z$ and $M_{xy}$ do not radiate in the direction perpendicular to the metasurface plane, i.e., at $\mathbf{n}=(0, 0, \pm 1)$ the electric field components are $E_x = E_y = E_z = 0$. Therefore, they do not contribute to the transmitted and reflected far-fields under normal excitation. On the other hand, they can have a ``side'' radiation if the scattering direction is not perpendicular to the metasurface plane, i.e., at $n_x \neq 0$, $n_y \neq 0$ and $n_z \neq 0$. However, in the array, the ``side'' radiation of each dipole moment $p_z$ (or quadrupole moment $M_{xy}$) located at lattice points will be cancelled (destructive interference) in the far-field by the radiation of neighbouring dipoles (quadrupoles). This is illustrated schematically in Fig.~\ref{fig:m_bian}(e), where it is shown that the electric field of the dipole located at $l=0$ is fully cancelled by the electric field of the dipoles located at $l=-1$ and $l=1$. This can also be seen mathematically from Eqs.~(\ref{Esca2x}--\ref{Esca2y}). For any considered multipole radiation with positive $n_x$ (or $n_y$), one can find the negative one. Since $n_z$ is the same for all of them, the total $E_x$ and $E_y$ is zero in the far field. In Fig.~\ref{fig:m_bian}(e), we note that the total electric field ($E_z$ component) is not zero in the $xy$-plane (metasurface plane). Indeed, in the $xy$-plane $\mathbf{n}=(n_x, n_y, 0)$, from Eq.~(\ref{Esca2z}), one can see that $E_z \neq 0$. 
This field provides a contribution to the near field and can be associated (under certain parameters) with a non-radiative, symmetrically protected bound state in the continuum \cite{evlyukhin2021polarization}. The above analysis can also be done for the electric quadrupole moments $Q_{xx}$ and $Q_{yy}$.

It should be noted that the consideration of the case of an asymmetric unit cell once again clearly demonstrates the dependence of the multipole approximation on the point of localization of multipole moments in the metasurface. Indeed, a suboptimal choice of unit cell can lead to the appearance of multipole moments that do not directly contribute to the reflection and transmission coefficients, but are merely a consequence of the relative change in the electromagnetic field distribution within the selected cell.

\section{ DQM at arbitrary oblique incidence\label{sec:examples}}

\begin{figure}[!htb]
\centering
\includegraphics[width=0.7\linewidth]{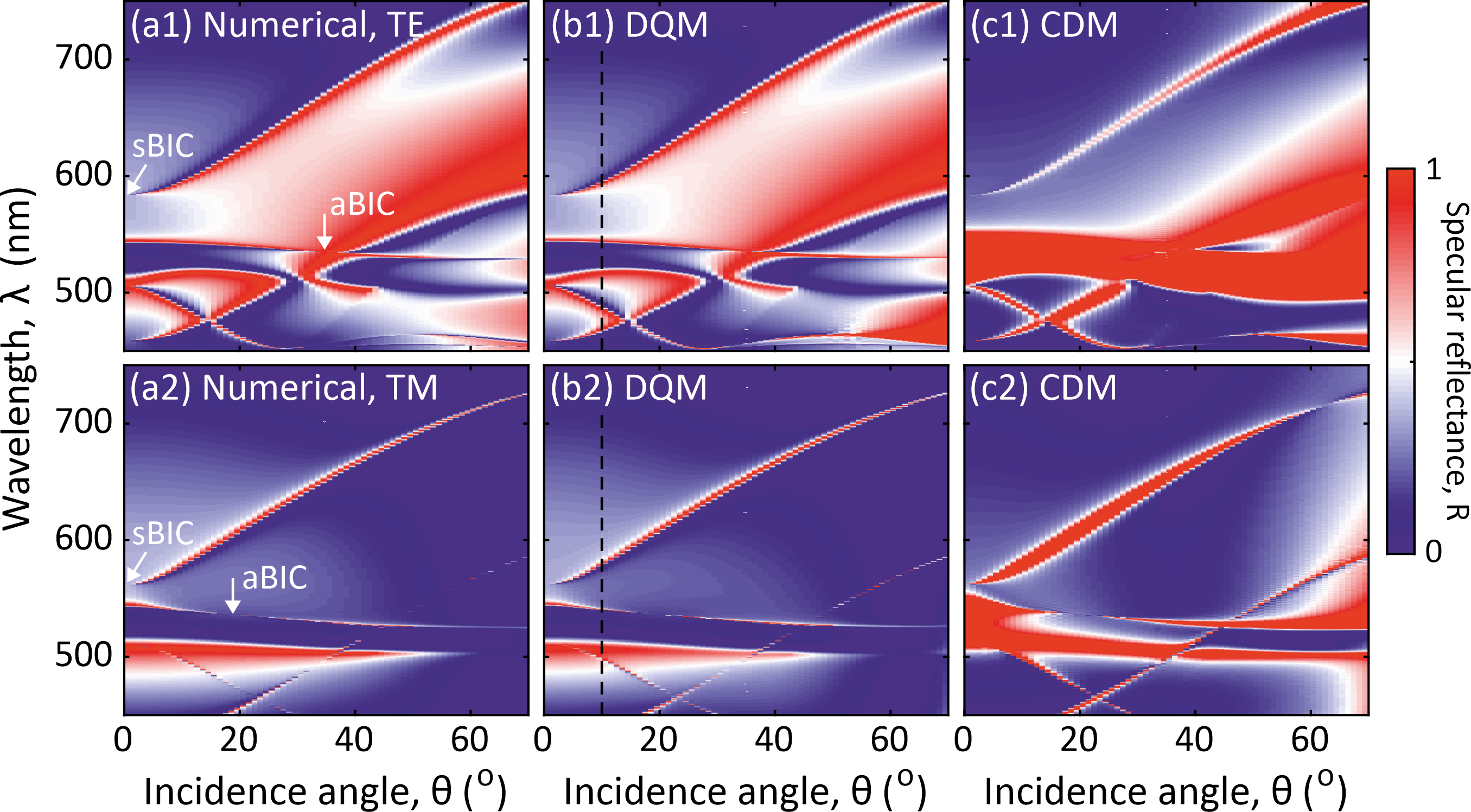}
\caption{Comparison of (a) full numerical, (b) our dipole-quadrupole model (DQM), and (c) coupled-dipole model (CDM) based specular reflectance of the structure considered in Fig.~\ref{fig:rmuc} as a function of the incidence angle $\theta$ at $\varphi=0$ for both (top panels, 1) TE and (bottom panels, 2) TM polarizations. In DQM and CDM calculations, the unit cell configuration UC1 is considered. sBIC: symmetry-protected BIC, aBIC: accidental BIC. The vertical dashed lines show the position of the sections shown in Fig.~\ref{fig:Rt10}.}
\label{fig:rtuc1}
\end{figure}

In this subsection, we discuss the applicability of our DQM. For demonstration, we calculate the angular dependence of the reflectance of the same membrane metasurface considered in Fig.~\ref{fig:rmuc} for both TE and TM polarized incidence. The unit cell configuration UC1 is used for dipole and quadrupole moments calculations. The results presented in Fig.~\ref{fig:rtuc1} show that DQM has a very good agreement with full numerical simulation for both polarizations. The model fully reproduced all spectral features, including very narrow ones, in a wide range of incident angles. Furthermore, in Fig.~\ref{fig:rtuc1}~(c1) and (c2), we see that without quadrupole contributions, i.e., within only the dipole approximation (CDM), it is impossible to obtain agreement with precise numerical calculations.

\begin{figure}[!htb]
\centering
\includegraphics[width=0.5\linewidth]{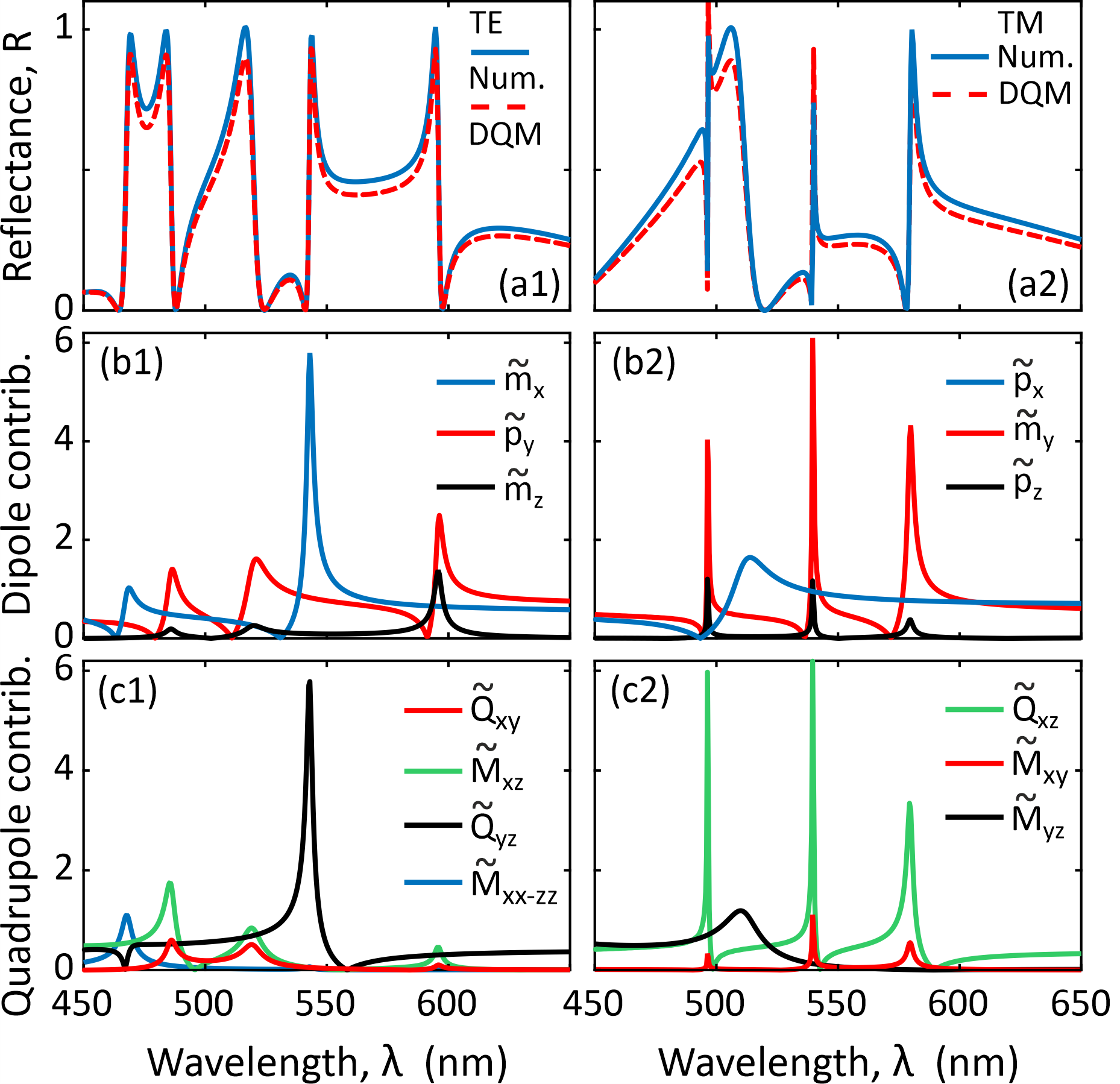}
\caption{Multipole decomposition of (a) metasurface resonances (red dashed: numerical, blue solid: DQM) into (b) dipole and (c) quadrupole contributions for (1) TE and (2) TM polarized illumination at $\theta=10^{\circ}$ and $\varphi=0$. The absolute dipole and quadrupole contributions [moments with tilde in (b1--c2)] are plotted in a unitless form, i.e., they include all coefficients (including angular dependencies) of the moments in Eqs.~\eqref{rTE} and ~\eqref{rTM}. In (c1), $\tilde{M}_{xx-zz} = \tilde{M}_{xx} - \tilde{M}_{zz} = 2(\tilde{M}_{xx} + \tilde{M}_{yy}/2)$. 
The metasurface parameters are as in Fig.~\ref{fig:rtuc1}.}
\label{fig:Rt10}
\end{figure}

Now, let us consider a single angle of incidence and use DQM to demonstrate a multipole analysis of reflectance resonant features. Figure~\ref{fig:Rt10} shows the reflectance from Fig.~\ref{fig:rtuc1} at fixed $\theta=10^{\circ}$ and its dipole and quadrupole contributions for both polarizations. 
We clearly identify the multipole origin of each reflectance peak. For instance, the resonance occurring around $\lambda=540$~nm in the reflection spectrum in the TE case is mostly due to contributions of resonant MD $m_1^{\parallel}=m_x$ and EQ $Q_3^{\perp} = Q_{yz}$ moments. The excitation of MD and EQ resonances at (around) the same wavelength indicates the existence of lattice coupling between these multipole moments. The contributions of other multipole moments, including the magnetic quadrupole, to this resonance are negligibly small. We note that these multipole resonances have a Fano-type (asymmetric) shape with zeros on the different sides of the resonance peak, and their amplitudes are comparable. Hence, one can expect that at a certain angle their zeros coincide, the resonance collapses, and the radiating resonant state becomes a nonradiating eigenstate of the metasurface, also known as aBIC~\cite{allayarov2024anapole}. Indeed, as shown in Fig.~\ref{fig:rtuc1}(a1), this happens at $\theta=35^{\circ}$ and at a slightly blueshifted wavelength of $\lambda=535$~nm. 

We observe a similar behavior in the TM case around $\lambda=540$~nm as well. The difference is that the resonance at $\lambda=540$~nm is quite narrow, and it occurs as a contribution of other four multipole moments: MD $m_2^{\perp}=m_y$, ED $p_3=p_z$, and EQ $Q_3^{\parallel} = Q_{xz}$ and MQ $M_1^{\perp} = M_{xy}$ moments. This in turn confirms the presence of coupling between these multipole moments in the metasurface under oblique  illumination. In this case, aBIC is realized at $\theta=20^{\circ}$ and $\lambda=535$~nm [Fig.~\ref{fig:rtuc1}(a2)].

\begin{figure}[!b]
\centering
\includegraphics[width=0.7\linewidth]{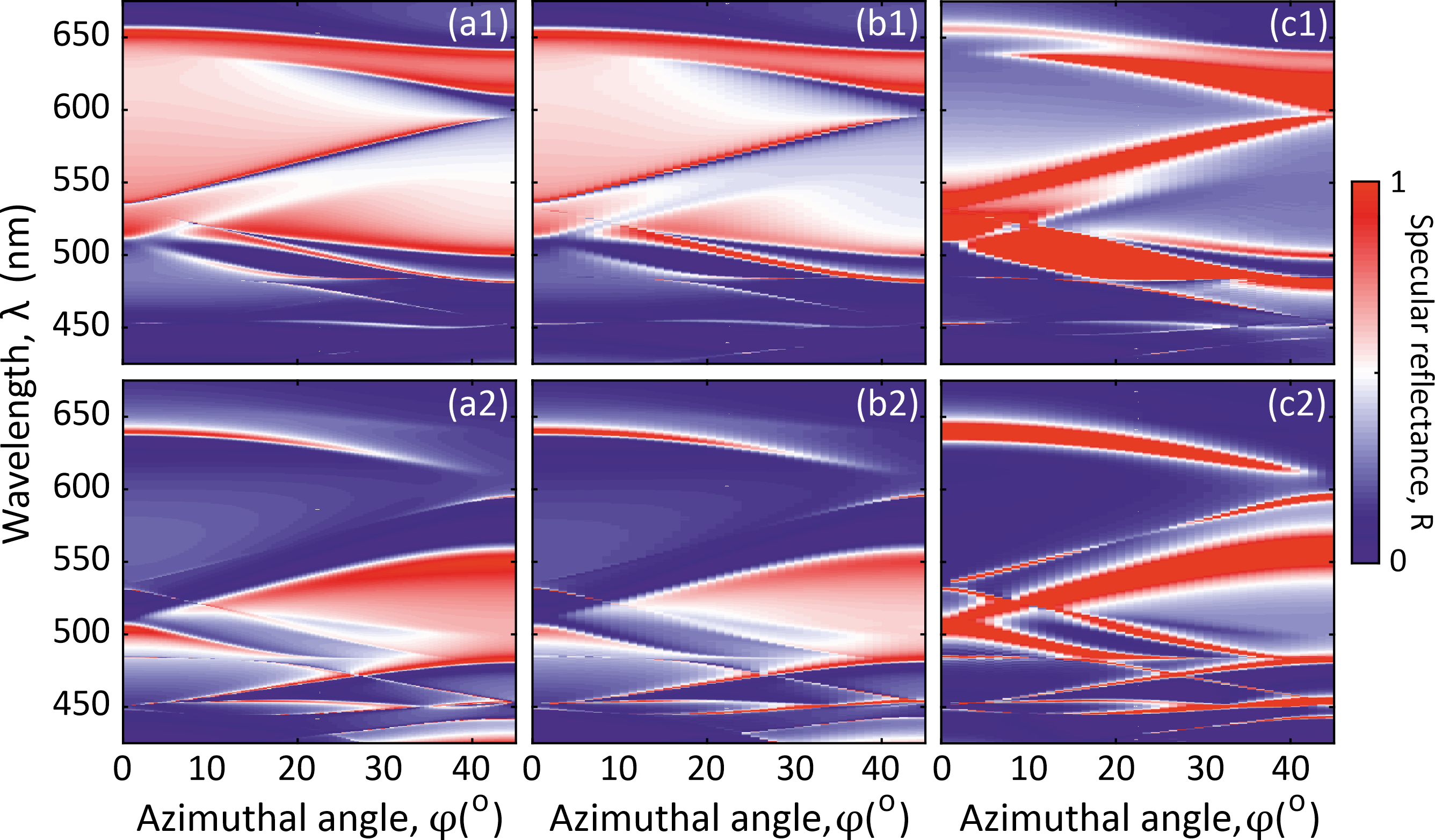}
\caption{Comparison of (a) full numerical, (b) our dipole-quadrupole model (DQM), and (c) coupled dipole model (CDM) based specular reflectance of the structure considered in Fig.~\ref{fig:rmuc} as a function of the azimuthal angle $\varphi$ at fixed $\theta=30^\circ$ for both (top panels, 1) TE and (bottom panels, 2) TM polarizations. In DQM and CDM calculations, the unit cell configuration UC1 is considered.}
\label{fig:rtuc1phi}
\end{figure}

A comparison of the Fano resonances in Fig.~\ref{fig:Rt10}(a1,a2) around  $\lambda=580$~nm  shows a different hierarchy of the spectral position of their maximum and minimum: for TE and TM polarization, the maximum is located on the short- and long-wavelength side, respectively. Such different behavior can be associated with different multipole contributions to these resonances. For TE polarization, the main (dominant) contribution is determined only by the dipole terms, see Fig.~\ref{fig:Rt10}~(b1) and (c1) around $\lambda=580$~nm, so  we can write using Eq.~\eqref{rTE} 
\begin{equation}\label{TE_2}
    r^{\rm TE}\sim\left(\left[p_y+\frac{1}{v}m_z\sin\theta\right]+\frac{1}{v}m_x\cos\theta\right).
\end{equation}
In this case, the resonant contributions $m_z$ and $p_y$ mainly arise from the excitation of symmetry-protected quasi-BIC resonances associated with $m_z$ and its coupling to the SLR associated with $p_y$ (the occurrence of such coupling was discussed in~\cite{allayarov2025analytical}). Thus, the suppression of reflection on the long-wavelength side of the Fano resonance is determined by the overlap of the hybrid resonance of $m_z$ and $p_y$ with the non-resonant contribution $m_x$, so that $r^{\rm TE}\to 0$ in (\ref{TE_2}). The excitation of quadrupole moments does not play a significant role here.
In the case of TM polarization, see Fig.~\ref{fig:Rt10}~(b2) and (c2), both dipole and quadrupole terms contribute to the Fano resonance, so that
\begin{equation}\label{TM_2}
    r^{\rm TM}\sim\left(\left[\frac{1}{v}m_y - p_z\sin\theta - \frac{{\rm i}k_{\rm s}}{6} Q_{xz} \cos2\theta - \frac{{\rm i}k_{\rm s}}{2v} M_{xy} \sin\theta\right]-p_x\cos\theta\right).
\end{equation}

Thus, the Fano resonance profile is determined by the resonant contributions of the dipole and quadrupole moments in square brackets of Eq.~(\ref{TM_2}) and the non-resonant contribution $p_x$. The simultaneous  resonant excitation of the multipoles from square brackets indicates their mutual electromagnetic coupling in the metasurface under oblique illumination with $\varphi=0$. In this hybrid resonance, $p_z$ and $M_{xy}$ originate from the symmetry-protected quasi-BIC, and $Q_{xz}$ and $m_y$ from the SLR. The change in the multipole composition of the Fano resonance for TM polarization compared to that of TE polarization leads to a difference  in the spectral hierarchy of  the maximum and minimum of the reflectance shown in Fig.~\ref{fig:Rt10} around $\lambda=580$~nm.

Finally, note that similar comparison and multipole analysis of resonances, as presented above, can also be done for a varying azimuthal angle $\varphi$. As an example, in Fig.~\ref{fig:rtuc1phi}, the above comparison is shown for a varying azimuthal angle $\varphi$ (rotation angle of the plane of incidence) at the fixed incidence angle of $\theta=30^\circ$. As one can see, in this case also our DQM shows very good agreement with the full numerical results, which cannot be obtained within only dipole approach.

\section{Application examples\label{sec:cases}}

In this section, we apply the multipole expansion presented and discussed above to analyze a number of resonant features arising in membrane metasurfaces.

\subsection{``Anti-Fano'' or anapole-assistant Fano resonance\label{sec:case1}}

Fano resonances in optical systems typically occur when a narrowband resonant radiation channel overlaps with a broadband background radiation. In this case, the maximum radiation intensity is determined by the contribution of the resonant radiation channel. When the narrowband region of resonant suppression of any radiation channel overlaps with broadband background radiation, we can consider this as an ``anti-Fano" resonance, since in this case the peak of  radiation intensity is determined basically by the contribution of the background radiation. In this paper, we consider this effect for membrane metasurfaces.

\begin{figure}[!t]
\centering
\includegraphics[width=0.8\linewidth]{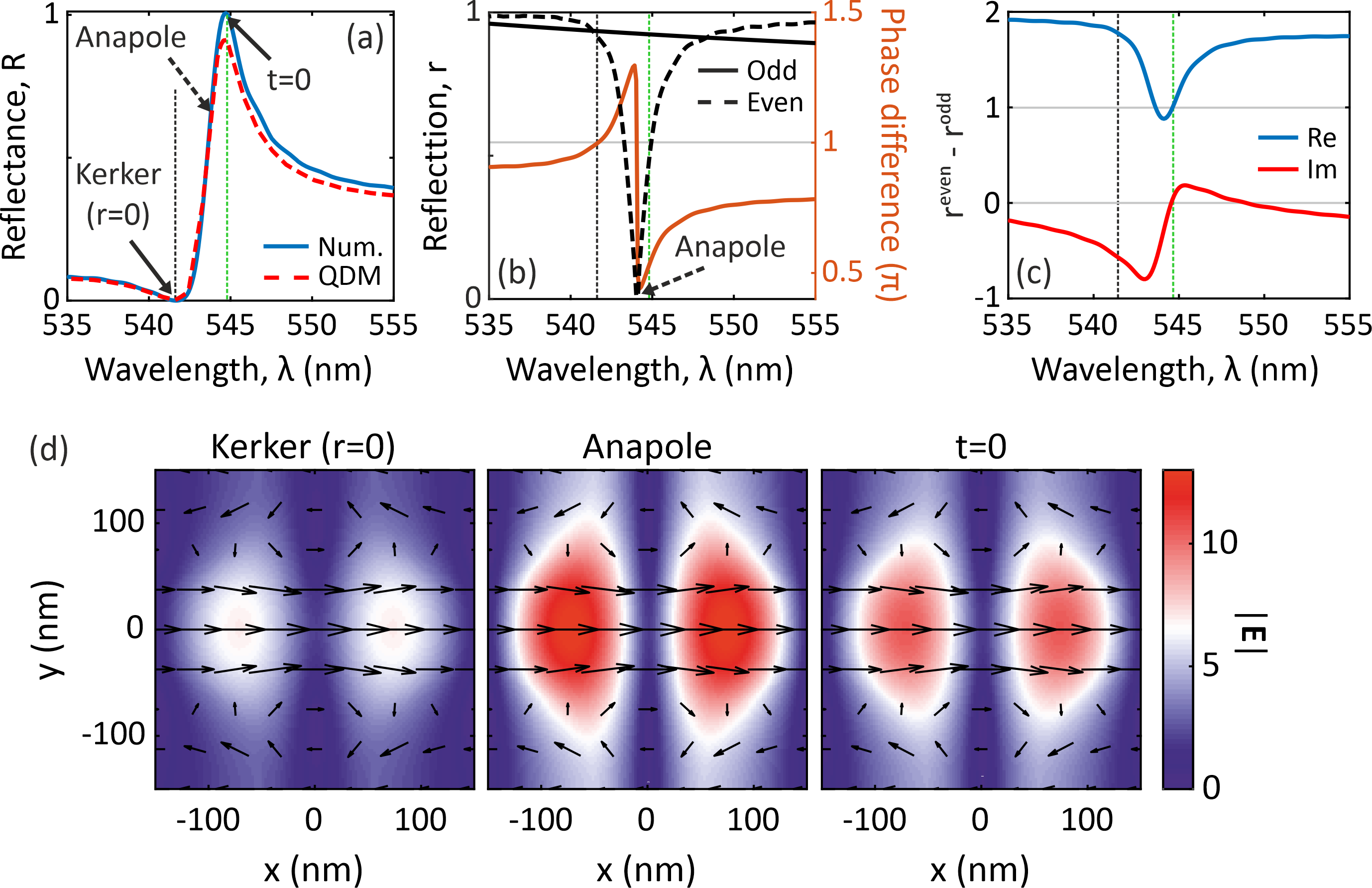}
\caption{(a) Reflectance of the membrane metasurface considered in Fig.~\ref{fig:rmuc}(a1) under the normal excitation. Kerker: lattice Kerker effect (suppression of reflection $r=0$); $t=0$: suppression of transmission. (b) Odd ($|r^{\rm odd}|$) and even ($|r^{\rm even}|$) multipole contributions to the total reflectance and the phase difference $\phi^{\rm odd}-\phi^{\rm even}$ between them. (c) Real and imaginary parts of the difference between $r^{\rm odd}$ and $r^{\rm even}$. (d) Electric field distribution and direction (the black arrows)  on the $xy$-plane (z=0) and at the spectral points indicated by arrows in (a).}
\label{fig:rmuc1}
\end{figure}

Let us analyze in detail the narrow resonance at $\lambda=545$~nm shown in Fig.~\ref{fig:rmuc}(a1). By looking at its multipole decomposition in Fig.~\ref{fig:rmuc}(b1), it seems that the resonant peak at $\lambda=545$~nm  in the reflectance is due to the strong MD and EQ resonances appearing at the same wavelength. Below we show that the origin of the resonant maximum in reflectance is contrarily due to contribution of the non-resonant and relatively weak multipole moments such as ED and MQ. For our further analysis, we use the effective dipole approach~\cite{allayarov2024multiresonant,allayarov2024anapole}. Within this approach the total reflection coefficient $r$ in Eq.~(\ref{r16}) can be divided into two independent contributions of  odd (e.g., ED, MQ) and even (e.g., MD, EQ) multipole subsystems, i.e., $r=r^{\rm odd}+r^{\rm even}$ and the transmission coefficient as $t=1+r^{\rm odd}-r^{\rm even}$. For the sake of convenience, Fig.~\ref{fig:rmuc1}(a) shows the same resonance in a narrower spectral range. 
Such resonance has a minimum and a maximum, which correspond to the lattice Kerker effect and total reflection, respectively. As shown in Fig.~\ref{fig:rmuc1}(b), the contribution of the even (dominating and resonant MD and EQ) multipoles $r^{\rm even}$ to $r$ vanishes at the resonance. This means that they interfere destructively in the far-field region and form a nonradiating anapole state. We call it even-anapole. Resonance in the reflection occurs due to the different interference mechanisms of non-resonant odd multipoles and with even multipoles on both sides (spectrally) of the even-anapole. On its shorter wavelength side, the Kerker effect takes place, i.e., destructive interference between odd and even multipoles in the backward direction. Indeed, in Fig.~\ref{fig:rmuc1}(b) we observe that where $r=0$ (black vertical dashed line), $|r^{\rm odd}|=|r^{\rm even}|$ and $\phi^{\rm odd}-\phi^{\rm even} = \pi$. One can also show that $r=r^{\rm odd}+r^{\rm even}=|r^{\rm odd}|\,\mathrm{exp} (\mathrm{i} \phi^{\rm odd}) [1+\mathrm{exp}(\mathrm{i}\pi)]=0$. On the longer wavelength side of the even-anapole, at a certain wavelength, $t = 0$. The physical mechanism behind it is the destructive interference between the incident wave and forward-scattered odd and even multipole radiations. Mathematically it means $r^{\rm even} - r^{\rm odd} = 1$. Indeed, in Fig.~\ref{fig:rmuc1}(c), we see that at $t=0$ (green vertical dashed line) the condition is satisfied. 
Thus, from Fig.~~\ref{fig:rmuc1}, we deduce that the main contribution to the reflection maximum comes from the non-resonant subsystem of odd multipoles, while the contribution of resonant even multipoles is significantly suppressed due to the anapole effect. Such behavior can be attributed to anti-Fano reflection resonance.

The anapole state is associated with near-field enhancement. Accordingly, it is expected that the near-field amplitude will be maximal exactly at the even-anapole position, and not at the lattice Kerker or $t=0$ position, as also confirmed in Fig.~\ref{fig:rmuc1}(d).

\subsection{Incidence angle independent resonance\label{sec:case2}}

\begin{figure}[!t]
\centering
\includegraphics[width=0.95\linewidth]{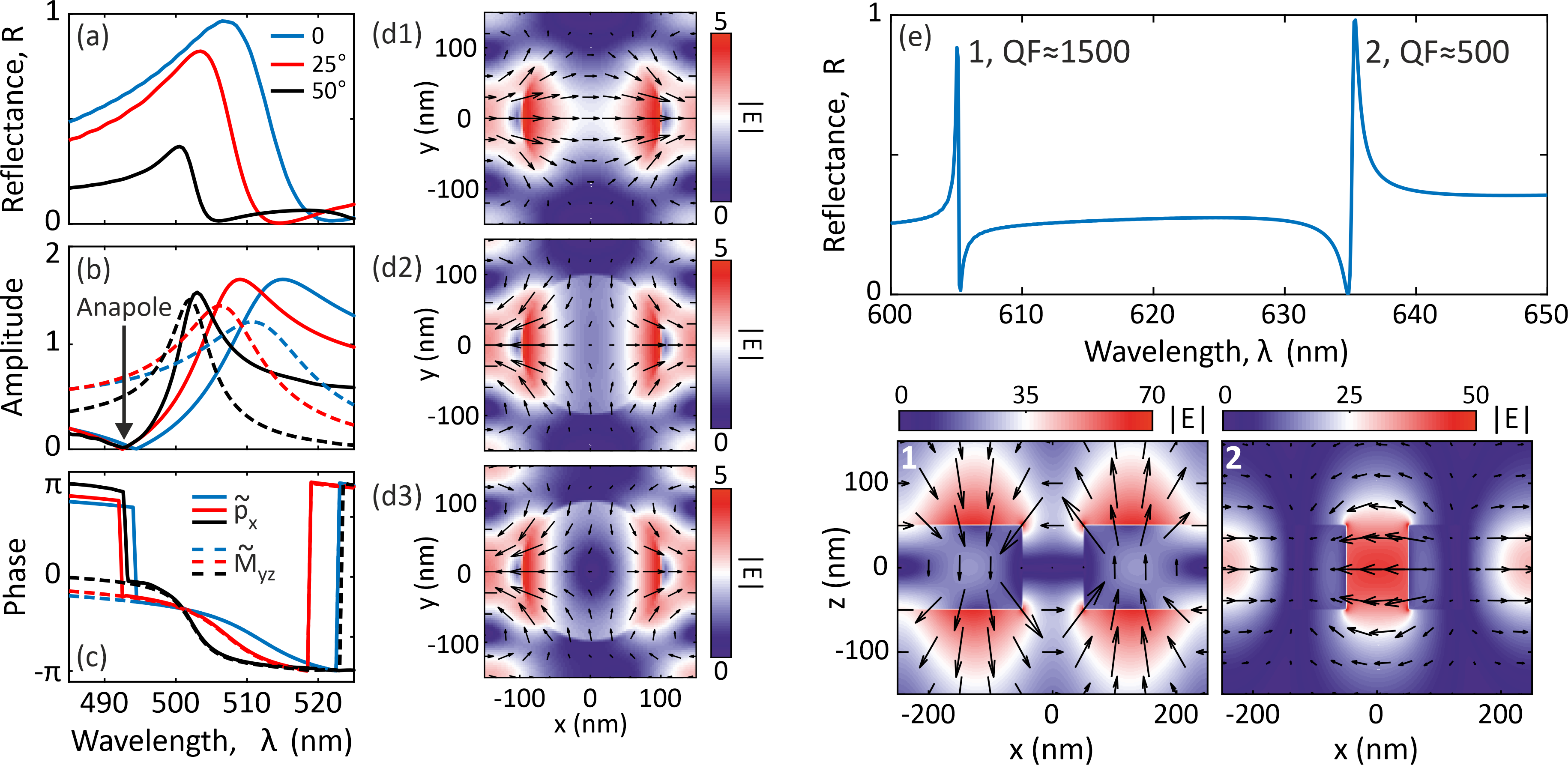}
\caption{(a) Reflectance (DQM for TM polarization), (b) normalized amplitudes and (c) phases of the in-plane parallel ED ($\tilde{p}_x$) and out-of-plane perpendicular MQ ($\tilde{M}_{yz}$) contributions for $\theta = [0,25^{\circ},50^{\circ}]$. The metasurface parameters are as in the case of Fig.~\ref{fig:rtuc1}. (d) Electric field distribution and direction ( the black arrows) on the $xy$-plane ($z=0$) at 1) $\theta = 0$, 2) $\theta = 25^{\circ}$, 3) $\theta = 50^{\circ}$, and $\lambda=500$~nm. (e) Reflectance of the membrane-metasurface shown in Fig.~\ref{fig:str} for $H=100$~nm, $D=100$~nm, $P=500$~nm and the normal incidence $x$-polarized plane wave excitation. Bottom panels depict the electric field distributions and directions (the black arrows)  on the $xz$-plane ($y=0$) at the positions 1 and 2 ($\lambda_1 = 605$~nm and $\lambda_2 = 635$~nm, respectively). The fields are normalized to the incident field. QF: quality factor.}
\label{fig:TMField}
\end{figure}

Another interesting behavior observed in Fig.~\ref{fig:rtuc1}(a2) for TM polarization is that the relatively broad resonance occurring at $\theta=0$ and $\lambda=500$~nm practically does not depend on the incidence angle ($\approx 5$~nm blueshift of the resonance peak in the $\theta \in [0, 60^\circ]$ range) and becomes narrower by increasing the angle $\theta$. To explain this behavior, let us consider that resonance at three different angles, $\theta=[0,25^{\circ},50^{\circ}]$ (see Fig.~\ref{fig:TMField}(a)), and decompose into multipole contributions [amplitude and phase, Fig.~\ref{fig:TMField}~(b) and (c)]. 
From the multipole decomposition, we see that the resonance is a result of the superposition of the in-plane parallel ED ($p_1^{\parallel}=p_x$) and out-of-plane perpendicular MQ ($M_3^{\perp} = M_{yz}$) components. 
Furthermore, they are coupled [see Eq.~(\ref{rTM})], therefore, the multipole contributions in the reflection remain in-phase for the considered angles of incidence. Both the ED and MQ resonances blueshift slightly and get narrower when the angle $\theta$ is increased. In Fig.~\ref{fig:TMField}(b), it is shown that the zero of ED resonance (anapole) is almost not shifted. Hence, one can say that the ED resonance can shift only up to this zero over $\theta$. 
This is true for MQ resonance, since it is coupled to ED resonance. Due to these behaviors, the reflectance experiences only a small resonance shift over a wide range of angles of incidence. A similar feature, but for a narrower resonance, is also present in the case of TE polarization [resonance around $\lambda=550$~nm in Fig.~\ref{fig:rtuc1}(a1)]. 
These features can have a potential application, e.g., in sensing and nonlinear optics, since the electric field is mostly concentrated in an accessible air hole region and its amplitude, enhanced ca. 5 times, remains also stable (see Fig.~\ref{fig:TMField}~(d1)-(d3)). 

Here, it is important to note that by optimizing the membrane-metasurface design (e.g., its thickness, hole diameter and period), one order of magnitude in electric field enhancement can be gained inside the hole as well as at the surface of the membrane. Indeed, as shown in Fig.~\ref{fig:TMField}(e), for the optimized metasurface with thickness $H=100$~nm, hole diameter $D=100$~nm, and period $P=500$~nm under $x$-polarized plane wave excitation, the electric field is 50 and 70 times enhanced in the hole and at the surface of the membrane, respectively.

\subsection{Bianisotropic membrane metasurfaces\label{sec:case3}}

\begin{figure}[!b]
\centering
\includegraphics[width=1\linewidth]{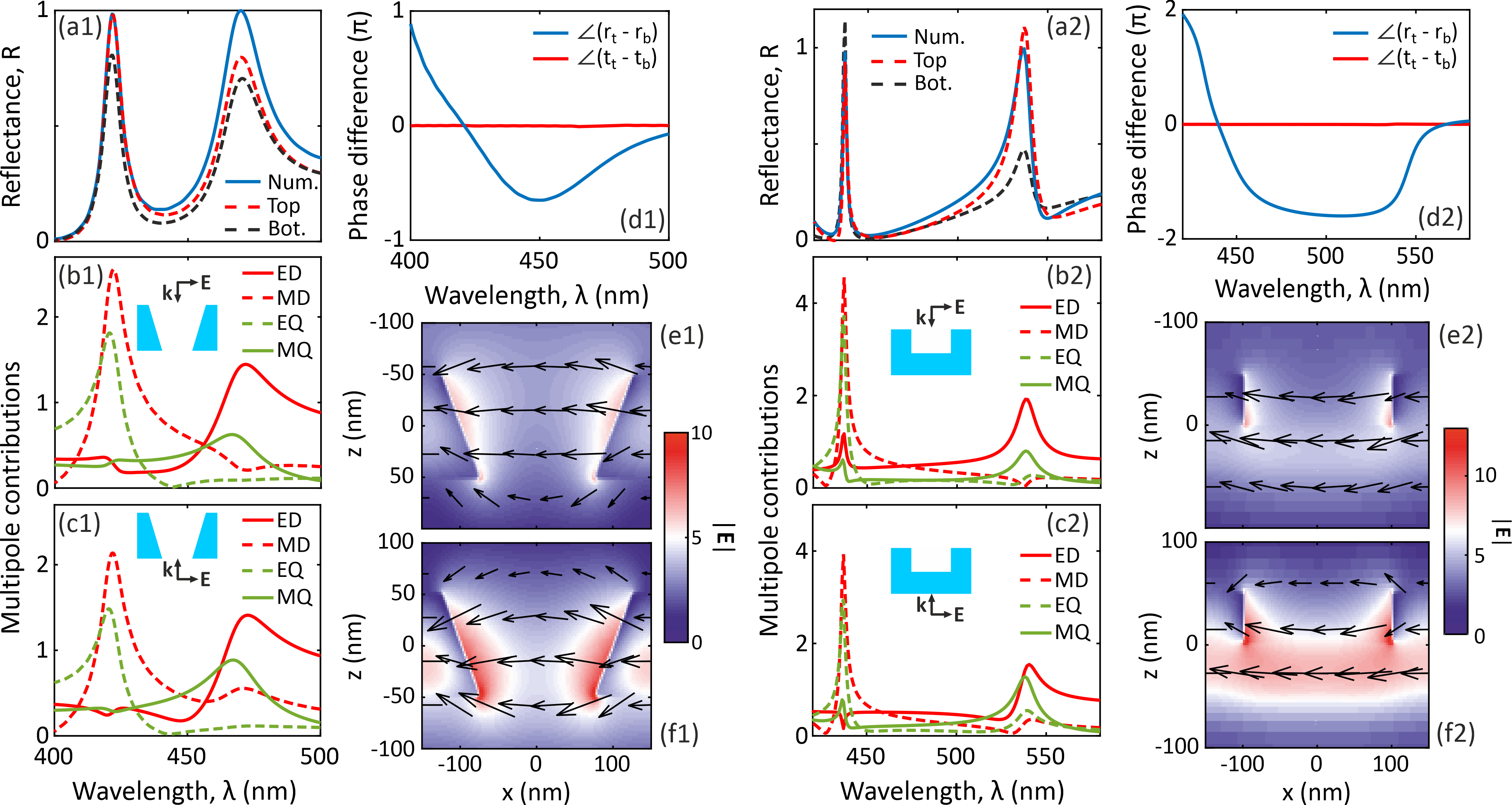}
\caption{(a1) Numerical (solid line) and QDM (dashed lines) based reflectance for top [see inset of (b1)] and bottom [see inset of (c1)] normal excitations and their multipole contributions, shown in (b1) and (c1), respectively. The structure is a conical hole membrane metasurface with $D_{\rm t} = 250$~nm and $D_{\rm b} = 150$~nm. (d1) Phase difference of reflection and transmission coefficients between top (t) and bottom (b) excitations. Electric field distribution and direction (the black arrows) on the $xz$-plane ($y=0$) under (e1) top and (f1) bottom excitation at $\lambda=470$~nm. The panels (a2-f2) show the same results for a half-perforated membrane metasurface. In (e2,f2), the considered wavelength is $\lambda=540$~nm. The unit cell configuration UC1 is considered for the multipole decomposition. Period and thickness are $P=300$~nm and $H=100$~nm, respectively, in both cases.}
\label{fig:conical}
\end{figure}

In the above sections, as an example, we have considered a membrane metasurface with fully perforated cylindrical holes and demonstrated that the presented approach works very well. In this section, we consider an array of more complex holes, namely conical and partially perforated holes, which may be relevant to account for fabrication constraints. 
Such membrane metasurfaces are particularly interesting because they exhibit bianisotropy, due to the fact that their unit cells will always have non-inversion symmetry regardless of the choice of cell's position \cite{poleva2023multipolar}. 
Further, to make the analysis simpler, we consider normal incidence excitation. However, it can be applied to an arbitrary angle of incidence. Let us first discuss the membrane metasurface with conical holes and its far- and near-field properties under top (incidence direction along the $z$-axis) and bottom (incidence direction opposite the $z$-axis) excitations. 
Due to the reciprocity principle, reflectance and transmittance do not depend on the excitation direction~\cite{shelankov1992reciprocity,potton2004reciprocity}. In Fig.~\ref{fig:conical}(a1), the reflectance calculated based on the full numerical simulations (solid line) and DQM (dashed lines) is presented. The dipole-quadrupole approach shows good agreement with the numerical result for both excitation directions. 
The difference between the top and bottom excitation results is due to the change of the multipole contributions [see panels (b1) and (c1)] {which is caused by the bianisotropic coupling between multipole moments of different parities \cite{poleva2023multipolar}}. 
For example, the amplitude of the magnetic dipole contribution (at $\lambda=420$~nm) in the case of top excitation is approximately 1.5 times higher compared to the bottom excitation one. Moreover, from the multipole contributions shown in Fig.~\ref{fig:conical}~(b1) and (c1), it is evident that in the resonance region, the multipoles of both parities have spectral features caused precisely by their bianisotropic coupling. 
Another interesting difference is the phase of the reflected and transmitted waves. In the case of an array of bianisotropic lossless dielectric nanoparticles, it has been shown that when the excitation direction changes from top to bottom, the phase of the reflected wave changes while the transmitted one does not~\cite{alaee2015all}. 
In Fig.~\ref{fig:conical}(d1), we observe that it holds for bianisotropic dielectric membrane metasurfaces as well. All these changes in the far-field properties are due to the change in the near-field. 
In Fig.~\ref{fig:conical}~(e1) and (f1), we show the electric field distributions on the $xz$-plane at $\lambda=470$~nm [the short-wavelength resonance in Fig.~\ref{fig:conical}(a1)], for the top and bottom excitations, respectively. Although the phase and distribution of the electric field are similar in both cases, we clearly see a relatively significant difference in the field amplitude. 

Similar results for the membrane metasurface consisting of a periodic half-perforated cylindrical holes are presented in Fig.~\ref{fig:conical}~(a2)-(f2). 
In Fig.~\ref{fig:conical}(a2), for the resonance at $\lambda=540$~nm, we note that considering multipoles up to quadrupoles may not be enough to have a full qualitative agreement between full numerical (blue solid line) and DQM (black dashed line) reflectance. 
This is because the non-perforated part of the layer requires a very large number of multipoles. However, the resonance feature is fully captured. In comparison to the conical holes case, the difference in the multipole contributions, the phase difference of the reflection, and near-fields are more pronounced. Especially large changes in the reflection phase can have important applications in designing efficient reflectarray systems~\cite{huang2007reflectarray,cheng2014wave}, while controlling electric field confinement and enhancement in holes via excitation direction is important in nonlinear optics and sensing.

\begin{figure}[!b]
\centering
\includegraphics[width=0.6\linewidth]{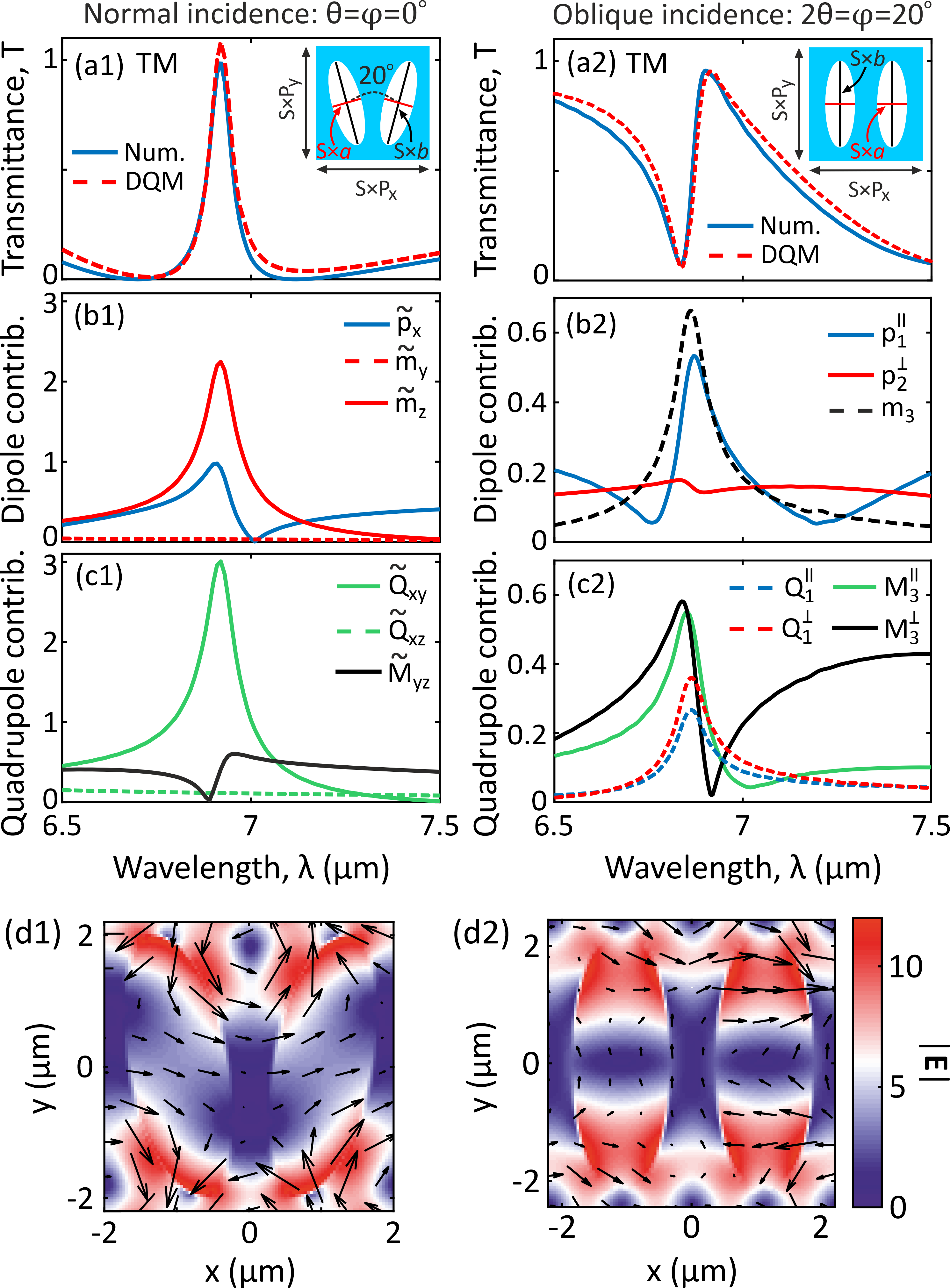}
\caption{(a) Reflectance and its (b) dipole and (c) quadrupole contributions (absolute values) of a membrane metasurface considered in Ref.~\cite{rosas2025enhanced}. (d) Electric field distribution on $xy$-plane ($z=0$) at $\lambda=6.9$~µm. (1) Elliptical holes with a $20^\circ$ tilt between their major axes [inset of (a1)]; normal incidence $x$-polarized excitation and scaling parameter $S=1$; (2) Parallel elliptical holes ($0^\circ$ tilt between their major axes) [inset of (a2)]; oblique incidence excitation with $\theta=10^\circ$, $\varphi=20^\circ$ and scaling parameter $S=1.11$. Insets in (a1,a2) show the considered unit cell configuration. The remaining parameters of the metasurface are $a = 1.4$~µm, $b = 4$~µm, $P_x = 4$~µm, $P_y = 4.4$~µm, $H = 1$~µm, $n_{\rm s} = 1$, $n_{\rm d} = 3.42$. }
\label{fig:optica}
\end{figure}

\subsection{Quasi-BIC under oblique incidence\label{sec:case4}}

As a final example, we analyze a membrane metasurface with a unit cell consisting of a pair of elliptical holes in a silicon membrane, which has been investigated in the mid-infrared region in Ref.~\cite{rosas2025enhanced}. In the paper, the authors excited quasi symmetry-protected BIC under normal incidence by tilting two parallel elliptical holes and obtained good agreement between numerical and experimental results. The multipole decomposition analysis of the resonance showed that MD and EQ have dominating contribution compared to ED and MQ ones~\cite{rosas2025enhanced}. To provide indirect experimental validation of our approach, we applied our dipole–quadrupole model to that structure. In Fig.~\ref{fig:optica}~(a1)-(d1), its transmittance, absolute dipole and quadrupole contributions and the electric field distribution at the resonance wavelength are shown. The metasurface parameters and excitation condition are as in Ref.~\cite{rosas2025enhanced}. 
The numerically calculated transmittance shows a very good agreement with the one based on the dipole-quadrupole approach. In our case as well, the resonance occurs due to the dominating out-of-plane MD ($m_z$) and ``in-plane'' EQ ($Q_{xy}$) moments and weaker in-plane ED ($p_x$) and ``out-of-plane'' MQ ($M_{yz}$) moments [see Fig.~\ref{fig:optica}~(b1) and (c1)]. Additionally, from our approach [see Eq.~(\ref{rTM})] one can see that the main resonant MD ($m_z$) and EQ ($Q_{xy}$) moments do not enter the reflection (and transmission) coefficient and cannot contribute directly to the far-field radiation. They radiate in the plane of the metasurface. Nevertheless, they contribute to the far-field radiation (reflection and transmission) indirectly via ED ($p_x$) and MQ ($M_{yz}$) moments as they couple to them.

On the other hand, it is well-known that the quasi symmetry-protected BIC can be realized by illuminating the symmetric structure at an angle \cite{Abujetas2020coupled,allayarov2025analytical}. 
In Fig.~\ref{fig:optica}~(a2)-(d2), a similar resonance at the same wavelength is demonstrated for symmetric and parallel elliptical holes. The sizes of the unit cell and holes are slightly increased (roughly $10 \%$), while the incidence angle is the same as the tilt angle of each elliptical hole in Fig.~\ref{fig:optica}(a1). 
Under oblique excitation, in addition to $p_1^{\perp} = p_x$ and $M_{3}^{\perp} = M_{yz}$, several other multipole moments are excited, which can be seen from Fig.~\ref{fig:optica}~(b2)-(c2). The main difference is the presence of a resonant out-of-plane magnetic dipole moment ($m_3$) and ``in-plane'' electric quadrupole moments ($Q_{1}^{\perp}$ and $Q_{1}^{\parallel}$) which occur only for oblique excitation [they have angular dependencies of $\sin\theta$ and $\sin2\theta$, respectively; see Eqs.~(\ref{tTE_perp1})-(\ref{tTE_perp4})] and drive the quasi symmetry-protected BIC. Another important difference and advantage of the second case is the relatively simple tuning mechanism of the resonance by the illumination angle. This allows us to use the same membrane metasurface at different resonance wavelengths. Furthermore, the fabrication process can be simplified since there is no need to introduce a tilt angle or, as shown in the previous sections (e.g., see Fig.~\ref{fig:rtuc1}), one can use a single circular hole per unit cell.

\section{Conclusion \label{sec:concl}}

In conclusion, we have derived a semi-analytical dipole-quadrupole approach that can be applied to the multipole analysis of transmission and reflection resonances of both conventional and membrane metasurfaces excited at an arbitrary angle of incidence.
We have demonstrated that our coupled dipole-quadrupole approach results have very good agreement with full numerical calculations. Our method is valid in a diffractionless spectral range since it only accounts for the zero-order diffraction, and requires that octupoles and further higher-order poles are negligible. Accordingly, we have tested different unit cell configurations of membrane metasurface and provided guidance on how to choose an optimal unit cell, which contains mostly dipoles and quadrupoles. 
From the formulas for transmission and reflection coefficients, we have identified which dipole and quadrupole moments can couple at oblique incidence, as well as the conditions for symmetry-protected and accidental bound states in the continuum. Additionally, we have demonstrated several important spectral features of membrane-metasurfaces. For example, under certain conditions, a membrane-metasurface can have ``anti-Fano'' type resonance features and resonances with almost angle-independent spectral position and varying quality factor, which can be explained within our model. Furthermore, we show that in membrane metasurfaces, it is possible to achieve electric fields with very high intensity inside holes and at the surface of the membrane. These features can have potential applications in sensing and nonlinear optics. Our theoretical results can be applied to various types of perforated metasurfaces, e.g., bianisotropic membrane metasurfaces consisting of conical, partially perforated, and a pair of holes per unit cell. 
The last section of our article is devoted to systems that have already been experimentally demonstrated. The applicability of our approach to such systems demonstrated the practical significance of our method for explaining the multipole mechanism of their resonances. Moreover, it was clearly shown that excitation of resonances in the selected experimental structures can be achieved not by replacing the original metasurface with another with altered unit cell symmetry, but by simply changing the irradiation conditions of the original metasurface. Furthermore, our approach allows one to account for fabrication constraints, such as conical holes with different entrance and exit diameters. Such profiles can arise, for example, in electron-beam lithography due to reduced etching efficiency, and in focused ion beam milling as a result of material redeposition. Thus, we believe that the results of our work are of broad interest to the scientific community due to the universality of the multipole approach demonstrated here, and for their practical application in nanophotonics.

\medskip
\textbf{Acknowledgements} \par
This work was supported by the Deutsche Forschungsgemeinschaft (DFG, German Research Foundation) under Germany’s Excellence Strategy within the Cluster of Excellence PhoenixD (EXC 2122, Project ID 390833453). We acknowledge the central computing cluster operated by Leibniz University IT Services (LUIS), which is funded by the DFG (project number INST 187/742-1 FUGG).

\section*{Appendices \label{sec:app}}
\appendix

\section{Fields generated by electric quadrupole  \label{sec:appA}}
The electric field ${\bf E}$ generated at a coordinate ${\bf r}=(x,y,z)$ by a single EQ located at ${\bf r}_0$ is~\cite{babicheva2019analytical}:
\begin{align}
    &{\bf E}({\bf r},{\bf r}_0)=\frac{-k_0^2}{6\varepsilon_0}\int \hat{G}({\bf r},{\bf r}'){\bf b}({\bf r}',{\bf r}_0)d{\bf r}',\label{EEQ1}
\end{align}
where
\begin{align}\label{b}
    &{\bf b} = \hat{Q}\nabla\delta({\bf r}'-{\bf r}_0) \quad \text{or} \quad b_\gamma = \sum_\tau Q_{\gamma\tau}\frac{\partial}{\partial \tau'}\delta({\bf r}'-{\bf r}_0), \, \text{and} \,\, \gamma, \tau = (x,y,z).
\end{align}
Using the symmetry properties of the EQ tensor $\hat{Q}$, i.e, $Q_{xy}=Q_{yx}$, $Q_{xz}=Q_{zx}$, $Q_{yz}=Q_{zy}$ and $Q_{xx}+Q_{yy}+Q_{zz}=0$, Eqs.~\eqref{GF_homo} and~\eqref{ddelta}, and Bloch's theorem, we obtain the following total electric field of the EQs in a metasurface:
\begin{align}
    E_x({\bf r})= &\frac{-k_0^2}{6\varepsilon_0} \sum_{l=0}^\infty \Big[ Q_{xx} \Big(\frac{\partial}{\partial x} + \frac{1}{k_{\rm s}^2}\frac{\partial^3}{\partial x^3} - \frac{1}{k_{\rm s}^2}\frac{\partial^3}{\partial x \partial z^2} \Big) +Q_{xy} \Big(\frac{\partial}{\partial y} + \frac{2}{k_{\rm s}^2}\frac{\partial^3}{\partial x^2 \partial y} \Big) + Q_{xz} \Big(\frac{\partial}{\partial z} + \frac{2}{k_{\rm s}^2}\frac{\partial^3}{\partial x^2 \partial z} \Big) \nonumber\\
    &+Q_{yy} \Big(\frac{1}{k_{\rm s}^2}\frac{\partial^3}{\partial x \partial y^2} - \frac{1}{k_{\rm s}^2}\frac{\partial^3}{\partial x \partial z^2} \Big) + Q_{yz}\frac{2}{k_{\rm s}^2}\frac{\partial^3 }{\partial x \partial y \partial z}\Big] \phi({\bf r},{\bf r}_l){\rm e}^{{\rm i}\mathbf{k}^{\parallel}_{\rm s}\cdot \mathbf{r}_l^{\parallel}}, \label{EQx1}\\
    E_y({\bf r})=&\frac{-k_0^2}{6\varepsilon_0} \sum_{l=0}^\infty\Big[Q_{xx} \Big(\frac{1}{k_{\rm s}^2}\frac{\partial^3}{\partial x^2 \partial y} - \frac{1}{k_{\rm s}^2}\frac{\partial^3}{\partial y \partial z^2} \Big) +Q_{xy} \Big(\frac{\partial}{\partial x} + \frac{2}{k_{\rm s}^2}\frac{\partial^3}{\partial x \partial y^2} \Big) + Q_{xz}\frac{2}{k_{\rm s}^2}\frac{\partial^3 }{\partial x \partial y \partial z}  \nonumber\\
    &+Q_{yy} \Big(\frac{\partial}{\partial y} + \frac{1}{k_{\rm s}^2}\frac{\partial^3}{\partial y^3} - \frac{1}{k_{\rm s}^2}\frac{\partial^3}{\partial y \partial z^2}\Big) + Q_{yz} \Big(\frac{\partial}{\partial z} + \frac{2}{k_{\rm s}^2}\frac{\partial^3}{\partial y^2 \partial z} \Big)\Big] \phi({\bf r},{\bf r}_l){\rm e}^{{\rm i}\mathbf{k}^{\parallel}_{\rm s}\cdot \mathbf{r}_l^{\parallel}}, \label{EQy1}\\
    E_z({\bf r})=&\frac{-k_0^2}{6\varepsilon_0} \sum_{l=0}^\infty\Big[Q_{xx}\Big(-\frac{\partial}{\partial z} - \frac{1}{k_{\rm s}^2}\frac{\partial^3}{\partial z^3} + \frac{1}{k_{\rm s}^2}\frac{\partial^3}{\partial x^2 \partial z} \Big) + Q_{xy} \frac{2}{k_{\rm s}^2}\frac{\partial^3 }{\partial x \partial y \partial z} + Q_{xz} \Big(\frac{\partial}{\partial x} + \frac{2}{k_{\rm s}^2}\frac{\partial^3}{\partial x \partial z^2} \Big) \nonumber\\
    &+Q_{yy} \Big(-\frac{\partial}{\partial z} - \frac{1}{k_{\rm s}^2}\frac{\partial^3}{\partial z^3} + \frac{1}{k_{\rm s}^2}\frac{\partial^3}{\partial y^2 \partial z} \Big) + Q_{yz} \Big(\frac{\partial}{\partial y} + \frac{2}{k_{\rm s}^2}\frac{\partial^3}{\partial y \partial z^2} \Big) \Big] \phi({\bf r},{\bf r}_l){\rm e}^{{\rm i}\mathbf{k}^{\parallel}_{\rm s}\cdot \mathbf{r}_l^{\parallel}}. \label{EQz1}
\end{align}
In the above equations, we used the relation $\phi({\bf r},{\bf r}')=\phi(|{\bf r}-{\bf r}'|)$ and $\dfrac{\partial}{\partial x'} \phi({\bf r},{\bf r}')=-\dfrac{\partial}{\partial x} \phi({\bf r},{\bf r}')$. The latter is true for other first (odd) order partial derivatives. Finally, using Eq.~(\ref{dphi1L0}) we obtain the specular fields:
\begin{align}\label{EEQ22}
    E_{\parallel} = \frac{{\rm i}k_{\rm s} \:{{\rm e}^{{\rm i}(k_xx+k_yy \pm k_zz)}}}{2S_{\rm L}\varepsilon_0\varepsilon_{\rm s}\cos\theta}\Big( \frac{{\rm i}k_{\rm s}}{6}\Big) \Big[& -\Big(\frac{3}{2}(Q_{xx}+Q_{yy})+\frac{1}{2}(Q_{xx}-Q_{yy})\cos2\varphi + Q_{xy}\sin2\varphi\Big)\sin\theta\cos^2\theta \nonumber\\
    &-(\pm)(Q_{xz}\cos\varphi + Q_{yz}\sin\varphi)\cos\theta\cos2\theta\Big], \\
    E_{\perp} = \frac{{\rm i}k_{\rm s} \:{{\rm e}^{{\rm i}(k_xx+k_yy \pm k_zz)}}}{2S_{\rm L}\varepsilon_0\varepsilon_{\rm s}\cos\theta}\Big( \frac{{\rm i}k_{\rm s}}{6}\Big) \Big[& -\Big(-\frac{1}{2}(Q_{xx}-Q_{yy})\sin2\varphi + Q_{xy}\cos2\varphi\Big)\sin\theta \nonumber\\
    &-(\pm)(-Q_{xz}\sin\varphi + Q_{yz}\cos\varphi)\cos\theta\Big],\\
    E_{z} = \frac{{\rm i}k_{\rm s} \:{{\rm e}^{{\rm i}(k_xx+k_yy \pm k_zz)}}}{2S_{\rm L}\varepsilon_0\varepsilon_{\rm s}\cos\theta}\Big( \frac{{\rm i}k_{\rm s}}{6}\Big) \Big[ & \Big(\frac{3}{2}(Q_{xx}+Q_{yy})+\frac{1}{2}(Q_{xx}-Q_{yy})\cos2\varphi + Q_{xy}\sin2\varphi\Big)\sin^2\theta(\pm\cos\theta) \nonumber\\
    &+(Q_{xz}\cos\varphi + Q_{yz}\sin\varphi)\sin\theta\cos2\theta\Big].
\end{align}
In the above equations, ``$-$" (``$+$") sign is for the reflected (transmitted) field. The magnetic field can be obtained from ${\bf H} = \dfrac{c\varepsilon_0}{{\rm i}k_0} \nabla \times {\bf E}$.

\section{Transmission coefficients  \label{sec:appB}}
The specular transmission coefficients for TE and TM polarized incidence can be obtained from the following definition:
\begin{align}
    &t^{\rm TE}_\perp=1+\frac{E_\perp^{\rm Total}}{E_{0\perp}{\rm e}^{{\rm i}{\bf k}_{\rm s}\cdot{\bf r}}}, \quad 
    t^{\rm TE}_\parallel=\frac{E_\parallel^{\rm Total}}{E_{0\perp}{\rm e}^{{\rm i}{\bf k}_{\rm s}\cdot{\bf r}}}, \\
    &t^{\rm TM}_\perp=1+\frac{H_\perp^{\rm Total}}{H_{0\perp}{\rm e}^{{\rm i}{\bf k}_{\rm s}\cdot{\bf r}}}, \quad 
    t^{\rm TM}_\parallel=\frac{H_\parallel^{\rm Total}}{H_{0\perp}{\rm e}^{{\rm i}{\bf k}_{\rm s}\cdot{\bf r}}},
\end{align}
where $E^{\rm Total}_{\parallel}$ ($H^{\rm Total}_{\parallel}$) and $E^{\rm Total}_{\perp}$ ($H^{\rm Total}_{\perp}$) are the parallel and perpendicular components of the total transmitted electric (magnetic) field of dipoles and quadrupoles for $z>0$, respectively. The transmittance is $T = |t_\perp|^2 + |t_\parallel|^2$.

Hence, by knowing the fields generated by dipole and quadrupole point sources, we obtain the following expression for the transmission coefficients:
\begin{align}
  t^{\rm TE}_\perp = &1 + \frac{{\rm i}k_{\rm s}}{2S_{\rm L}{E}_0\varepsilon_0\varepsilon_{\rm s}\cos\theta}\Big[p^\perp_2-\frac{1}{v}m^\parallel_1\cos\theta + \frac{1}{v}m_3\sin\theta \nonumber\\
  &+\frac{{\rm i}k_{\rm s}}{6}\Big(-Q_3^\perp\cos\theta - Q_1^\perp\sin\theta\Big) +\frac{{\rm i}k_{\rm s}}{2v}\Big(M_3^\parallel\cos 2\theta + \tilde{M}_1^\parallel\sin2\theta\Big)\Big], \label{tTE_perp1} \\
  t^{\rm TE}_\parallel = &\frac{{\rm i}k_{\rm s}}{2S_{\rm L}{E}_0\varepsilon_0\varepsilon_{\rm s}}\Big[p^\parallel_1\cos\theta-p_3\sin\theta+\frac{1}{v}m^\perp_2 \nonumber\\
  &+\frac{{\rm i}k_{\rm s}}{6}\Big(-Q_3^\parallel\cos2\theta - \tilde{Q}_1^\parallel\sin2\theta\Big) +\frac{{\rm i}k_{\rm s}}{2v}\Big(-M_3^\perp\cos \theta - M_1^\perp\sin\theta\Big)\Big], \label{tTE_perp2} \\
  t^{\rm TM}_\perp = &1 + \frac{t^{\rm TE}_\parallel}{\cos\theta},\label{tTE_perp3} \\
  t^{\rm TM}_\parallel = &\big(1 - t^{\rm TE}_\perp\big)\cos\theta. \label{tTE_perp4} 
\end{align}

In the case of $\varphi=0$, the transmission coefficients can be written as:
\begin{align}\label{tTE}
  t^{\rm TE} = 1 &+ \frac{{\rm i}k_{\rm s}}{2S_{\rm L}{E}_0\varepsilon_0\varepsilon_{\rm s}\cos\theta}\Big( \Big[ p_y + \frac{1}{v}m_z\sin\theta - \frac{{\rm i}k_{\rm s}}{6} Q_{xy} \sin\theta + \frac{{\rm i}k_{\rm s}}{2v} M_{xz} \cos2\theta \Big] \nonumber\\
  &+\Big[-\frac{1}{v}m_x\cos\theta - \frac{{\rm i}k_{\rm s}}{6} Q_{yz} \cos\theta + \frac{{\rm i}k_{\rm s}}{2v} (M_{xx} + \frac{1}{2} M_{yy})\sin2\theta \Big] \Big),
\end{align}

\begin{align}\label{tTM}
  t^{\rm TM} = 1 &+ \frac{{\rm i}k_{\rm s}}{2S_{\rm L}{E}_0\varepsilon_0\varepsilon_{\rm s}\cos\theta}\Big( \Big[ \frac{1}{v}m_y - p_z\sin\theta - \frac{{\rm i}k_{\rm s}}{6} Q_{xz} \cos2\theta - \frac{{\rm i}k_{\rm s}}{2v} M_{xy} \sin\theta \Big] \nonumber\\
  &+\Big[p_x\cos\theta - \frac{{\rm i}k_{\rm s}}{6} (Q_{xx} + \frac{1}{2} Q_{yy}) \sin2\theta - \frac{{\rm i}k_{\rm s}}{2v} M_{yz} \cos\theta \Big] \Big).
\end{align}

In Eqs.~\eqref{tTE} and \eqref{tTM}, the coupled multipole moments are grouped into square brackets. 

\section{Dipole and quadrupole moments \label{sec:appC}}
This Appendix contains expressions for the exact multipole moments that have been used in the reflection and transmission coefficients. They are listed here for the sake of completeness and for the reader's convenience. Originally, they were obtained in Refs.~\cite{alaee2018electromagnetic,evlyukhin2019multipole}.

Vector of the electric dipole (ED) moment:
\begin{align}\label{ed}
    {\bf p} = &\frac{{\rm i}}{\omega} \int_{V}j_0(k_{\rm s}r){\bf j}dV+\frac{{\rm i}k_{\rm s}^2}{2\omega} \int_{V} \frac{j_2(k_{\rm s}r)}{(k_{\rm s}r)^2} [3({\bf j} \cdot {\bf r}){\bf r}-{r}^2{\bf j}]dV.
\end{align}
Vector of the magnetic dipole (MD) moment:
\begin{align}\label{md}
    {\bf m} = &\frac{3}{2} \int_{V}\frac{j_1(k_{\rm s}r)}{k_{\rm s}r}({\bf r} \times {\bf j})dV.
\end{align}
Tensor of the electric quadrupole (EQ) moment:
\begin{align}\label{eq}
    {\hat{Q}} =& \frac{3{\rm i}}{\omega} \int_{V}\frac{j_1(k_{\rm s}r)}{k_{\rm s}r}[3({\bf r} {\bf j}+{\bf j} {\bf r})-2({\bf r} \cdot {\bf j})\hat{U}]dV \notag \\
    &+\frac{6{\rm i}k_{\rm s}^2}{\omega} \int_{V}\frac{j_3(k_{\rm s}r)}{(k_{\rm s}r)^3}[5({\bf r} \cdot {\bf j}){\bf r}{\bf r}-r^2({\bf j} {\bf r}+{\bf r} {\bf j}) - r^2({\bf r} \cdot {\bf j})\hat{U}]dV.
\end{align}
Tensor of the magnetic quadrupole (MQ) moment:
\begin{align}\label{mq}
    {\hat{M}} = &5 \int_{V}\frac{j_2(k_{\rm s}r)}{(k_{\rm s}r)^2}[({\bf r} \times {\bf j}){\bf r}+{\bf r}({\bf r} \times {\bf j})]dV.
\end{align}
where $r=|{\bf r}|$,  $j_n$ is the $n$-th order spherical Bessel function, ${\bf j}$ is the current density, $V$ is the volume of material (e.g., nanoparticle or membrane) in the unit cell. Note, in the above equations, it is assumed that the multipole moments are located at the origin of the coordinate system. If they are located at a point with radius vector ${\bf r}_0$, then $\bf r$ is replaced by ${\bf r}-{\bf r}_0$ in Eqs.~(\ref{ed}--\ref{mq}).  To calculate the multipole moments numerically, the total electric field ${\bf E}$ inside a nanoparticle has to be known. In the case of a membrane metasurface, one needs to use the total field in the portion of the unit cell occupied by the membrane's material. Then, the current density can be calculated as ${\bf j} = {-\rm i}\omega\varepsilon_0(\varepsilon_{\rm d}-\varepsilon_{\rm s}){\bf E}$, where $\varepsilon_{\rm d}$ is the relative permittivity of the nanoparticle or membrane's material. 

The total electric field ${\bf E}$ inside the nanoparticle can be numerically calculated, e.g., by using ANSYS Lumerical's FDTD solver with periodic boundary conditions along $x$ and $y$ directions, and perfectly matched layer (PML) boundary condition along $z$ direction. In the case of oblique incidence, it is necessary to use the broadband fixed angle source technique (BFAST) instead of the periodic boundary conditions~\cite{fdtd,liang2013wideband}.

\medskip

%
\bibliographystyle{MSP}
\bibliography{References}

\end{document}